\documentclass[letterpaper,11pt]{emulateapj}
\usepackage{epsfig}
\usepackage{natbib}
\usepackage{graphics}
\usepackage{graphicx}
\usepackage{multirow}
\usepackage{amssymb}
\usepackage{color}
\usepackage{boxedminipage}

\def \kms {{\rm km~s$^{-1}$}}
\def \kmsMpc {{\rm km~s$^{-1}$~Mpc$^{-1}$}}

\shorttitle{[3.6] TF Relation}
\shortauthors{Sorce et al.}

\begin{document}

\title{Calibration of the Mid-Infrared Tully-Fisher Relation}

\author{Jenny G. Sorce$^{1}$}
\email{j.sorce@ipnl.in2p3.fr}
\author{H\'el\`ene M. Courtois$^{1,2}$}
\author{R. Brent Tully$^{2}$}
\author{Mark Seibert$^{3}$}
\author{Victoria Scowcroft$^{3}$}
\author{Wendy L. Freedman$^{3}$}
\author{Barry F. Madore$^{3}$}
\author{S. Eric Persson$^{3}$}
\author{Andy Monson$^{3}$}
\author{Jane Rigby$^{4}$}
\affil{$^1$Universit\'e Claude Bernard Lyon I, Institut de Physique Nucleaire, Lyon, France} 
\affil{$^2$Institute for Astronomy, University of Hawaii, 2680 Woodlawn Drive, HI 96822, USA}
\affil{$^3$Carnegie Observatories, 813 Santa Barbara Street, Pasadena, CA 91101, USA}
\affil{$^4$Observational Cosmology Lab, NASA Goddard Space Flight Center, Greenbelt, MD 20771, USA}

\begin{abstract}


Distance measures on a coherent scale around the sky are required to address the outstanding cosmological problems of the Hubble Constant and of departures from the mean cosmic flow.  The correlation between galaxy luminosities and rotation rates can be used to determine distances to many thousands of galaxies in a wide range of environments potentially out to 200~Mpc.   Mid-infrared (3.6~$\mu$m) photometry with the Spitzer Space Telescope is particularly valuable as the source of the luminosities because it provides products of uniform quality across the sky.  From a perch above the atmosphere, essentially the total magnitude of targets can be registered in exposures of a few minutes.  Extinction is minimal and the flux is dominated by the light from old stars which is expected to correlate with the mass of the targets.

In spite of the superior photometry, the correlation between mid-infrared luminosities and rotation rates extracted from neutral hydrogen profiles is slightly degraded from the correlation found with $I$ band luminosities.  A color correction recovers a correlation that provides comparable accuracy to that available at $I$ band ($\sim20\%$ $1 \sigma$ in an individual distance) while retaining the advantages identified above.  Without the color correction the relation between linewidth and [3.6]  magnitudes is
$M^{b,i,k,a}_{[3.6]} = -20.34 - 9.74 ({\rm log} W_{mx}^{i} -2.5)$.   This description is found with a sample of 213 galaxies in 13 clusters that define the slope and 26 galaxies with Cepheid or tip of the red giant branch distances that define the zero point.  A color corrected parameter $M_{C_{[3.6]}}$ is constructed that has reduced scatter: $M_{C_{[3.6]}} = -20.34 - 9.13 ({\rm log} W_{mx}^{i} -2.5)$.  Consideration of the 7 calibration clusters beyond 50 Mpc, outside the domain of obvious peculiar velocities, provides a preliminary Hubble Constant estimate of H$_0=74\pm5$~\kmsMpc.\\ 

\end{abstract}
	


\section{Introduction}

Soon after the discovery of the power law correlation between the rotation rates of galaxies and their luminosities \citep{1977A&A....54..661T} it was suggested \citep{1979ApJ...229....1A} that the methodology might be improved by moving to near-infrared bands, particularly when it is used to measure distances.  Obscuration corrections within the hosts and due to our Galaxy are minimized and light from old stars, which peaks in the infrared, should optimally represent the baryonic mass that presumably couples to the rotation rate.  Progress with infrared observations of galaxies has been difficult, though, because of the high and variable sky foreground at near-infrared wavelengths and overwhelming thermal emission at mid-infrared wavelengths with ground-based observations.  The most serious modern attempts to use an infrared form of the correlation have drawn on the $K_s$ magnitudes of 2MASS, the Two Micron All-Sky Survey \citep{2002A&A...396..431K}.  However, these magnitudes, like with the earlier work in the infrared, only register the high surface brightness components of light from galaxies and can actually miss low surface brightness galaxies entirely.\\

The situation dramatically changed with the launch of {\it Spitzer Space Telescope} \citep{2004ApJS..154....1W}.  With observations using IRAC, the InfraRed Array Camera \citep{2004ApJS..154...10F}, the `sky' is far reduced from observations on the ground, now dominated by diffuse zodiacal light and the stochastic distribution of background high redshift galaxies.  Imaging with of order 4 minute integrations in the [3.6] band with this facility permits area photometry at levels that reach slightly fainter than ground-based optical imaging with comparable exposures; ie, to levels that include all but a few percent of the total light of a galaxy \citep{2012AJ....144..133S}.  In addition, and a very important point, the photometry has consistent properties in all directions on the sky.\\

Real progress on this program had to await the exhaustion of cryogenics on {\it Spitzer Space Telescope}.   During the subsequent `warm' mission, observations have only been possible with the two shortest wavelength passbands with the facility, at $3.6~\mu$m and $4.5~\mu$m, and there has been an emphasis toward large programs that can usefully work in these bands.  This article results from a commonality of interests between two of these programs.  One of these, initiated in Spitzer proposal cycle 6, is named {\it Carnegie Hubble Program} (CHP).   The intent of this program is to reduce systematics arising in the determination of the Hubble Constant.  A part of CHP gives attention to a mid-infrared calibration of the Cepheid Period-Luminosity relation and a second part addresses the properties of the rotation rate--luminosity correlation of galaxies, the Tully-Fisher Relation (TFR).  The two parts are related since the TFR zero point is established by the Cepheid distance measurements.  \citet{2011AJ....142..192F} describe the goals of CHP and \citet{freedman2012} report on the results of the Cepheid calibration that gives a distance modulus for the Large Magellanic Cloud of $18.48 \pm0.03$.  The second program, initiated in cycle 8, has the name {\it Cosmic Flows with Spitzer} (CFS).  The goals in this case are to acquire distances to several thousand galaxies using the mid-infrared TFR in order to map deviations from Hubble flow.  The two programs use overlapping data from Spitzer and require a similar calibration of the rotation rate - luminosity correlation.  This paper presents the calibration that will be used in subsequent work with both CHP and CFS.\\

The ensuing discussion borrows heavily on the recent re-calibration of the $I$ band correlation by \citet{2012ApJ...749...78T} (hereafter TC12).  That paper outlines a strategy of forming a template relation using samples from 13 galaxy clusters and the establishment of a zero point using nearby galaxies with independent Cepheid period-luminosity or Tip of the Red Giant Branch (TRGB) distances.   It turns out that [3.6] magnitudes now exist for a substantial majority of the same galaxies.  In this paper we use the same HI profile and inclination information as in the $I$ band calibration paper.  The only significant difference is the replacement of mid-infrared for optical luminosities.  It turns out that although the new photometry has high fidelity and the photometry correction terms are small there is an intrinsic color term in the [3.6] band TFR.  Scatter in the relation is reduced upon application of a color correction.  We conclude with an estimate of the Hubble Constant.


\section{Data}
\subsection{Calibrators}

The slope and zero point calibrator samples are described in detail in TC12.  The correlation slope is established from a template built from galaxies in 13 clusters.  The only departure in terms of an extension from the $I$ band calibration occurs in the case of Abell 2634.  The CHP program included observations of a larger region including Abell 2666.  The two clusters are close in projection and, evidently, in distance.  We find no discernable difference in distance between galaxies closest on the sky to A2634 versus those closest to A2666.  We propose to average over the entire complex. \\

Each cluster sample is comprised of galaxies likely to be at similar distances.  There was an attempt to include all galaxies with suitable properties down to a defined faint luminosity level to have an unbiased sampling of the cluster volume to a magnitude limit. Candidates are chosen out of a projection-velocity window. We care about minimizing relative distance effects in the TFR so it is more important to minimize interlopers than maximize true members.    Cluster members that are `window outsiders' would not be expected to lie in any preferred part of the TF diagram.  The selection criteria are:
 1) morphological types earlier than Sa are excluded,  2) HI profiles with adequate signal-to-noise are required (see next subsection), 3) no evidence of confusion or tidal disruption, 4) inclinations inferior to $45^{\circ}$ are rejected. Tests with samples that satisfy this limit have not revealed any distance bias with inferred inclinations (TC12).  Criteria for inclusion of zero point calibrators are similar, with the additional requirement that they have very well known distances from either Cepheid or TRGB measurements.  In our earlier papers, the Cepheid scale had been set by a distance modulus for the Large Magellanic Cloud of 18.50 \citep{2001ApJ...553...47F}.  Here we adopt the slightly modified modulus $18.48\pm0.03$ based on mid-infrared photometry of Cepheids in the LMC and in our Galaxy, the latter anchored with trigonometric parallaxes (Monson et al. 2012).  Our TRGB distances are based on a Population II calibration but have been demonstrated to be on a consistent scale \citep{2007ApJ...661..815R, 2008ApJ...676..184T}.\\	
 
With completion of the CFS program toward the end of 2012 the entire sample of calibrators used in the $I$ band calibration has been observed.  Because of the overlap in interests with CHP a large fraction of the $I$ band calibrators used by TC12 have already been observed in the earlier Spitzer cycle for the same purpose of a TFR calibration and most others have been observed serendipitously in other Spitzer programs.  At this time, 230 of 314 galaxies (73\%) used in the $I$ band calibration (plus 9 other galaxies introduced with the extension of the Abell 2634 sample to include Abell 2666) have Spitzer [3.6] photometry, including 26 of 36 (72\%) that set the zero point. 

 The completion is greater than 60\% with each of 12 of the 13 template clusters (the Pisces filament is the exception).  It is deemed appropriate to present a preliminary calibration with the available material.  In a later section there will be a review of the impact of the current completeness level on the small Malmquist bias that we make.  

\subsection{HI linewidths}

The Cosmic Flows project has now analyzed HI profiles for over 14,000 galaxies in a consistent way, deriving a linewidth parameter $W_{m50}$ with suitable precision (error estimate $\le 20$~\kms) for over 11,000 galaxies \citep{2009AJ....138.1938C,2011MNRAS.414.2005C}. This parameter is a measure of the HI profile width at $50\%$ of the mean flux within the velocity range encompassing $90\%$ of the total HI flux. The newly measured HI profiles of thousands of galaxies are available for public use at the Extragalactic Distance Database (EDD) website\footnote{http://edd.ifa.hawaii.edu; catalog `All Digital HI'}.  This observed parameter $W_{m50}$ is transformed into the more physically motivated parameter $W_{mx}^i$ through steps that are justified in \citet{2009AJ....138.1938C,2011MNRAS.414.2005C} and reviewed by TC12. $W_{mx}^i$ statistically approximates twice the maximum rotation velocity of a galaxy.\\

These transformations remove a slight relativistic broadening and a broadening due to finite spectral resolution, adjust to twice the projected maximum rotation velocity and de-project to edge-on orientation.  Linewidth error estimates are based on the level of the signal, $S$, at 50\% of mean flux divided by the noise, $N$, measured beyond the extremities of the signal. Profiles with error estimates smaller than 20 km s$^{-1}$ are retained. These profiles meet a minimum flux per channel requirement of $S/N \geq 2$ and acceptance after visual inspection.\\

Uncertainties in the rotation rate parameter are illustrated in the error bars of the figures presented in the next section.  It will be seen that errors in the linewidth parameter dominate observational uncertainties.  Errors in the logarithmic linewidth parameter tend to be larger for slow rotators since a typical measurement uncertainty of $10-20$ \kms\ causes a larger fractional uncertainty with a narrow profile.  The largest uncertainties are associated with more face-on galaxies, those toward the $45^{\circ}$ cutoff.  At this limit, a $5^{\circ}$ error in inclination results in an $8\%$ error in linewidth.  

\subsection{[3.6] Photometry}

The photometric data has all been obtained with IRAC ch.1, passband center $3.55~\mu$m.   CHP, the {\it Carnegie Hubble Program} \citep{2011AJ....142..192F}, provides 60\% of the data.  In addition, S$^{4}$G, the {\it Spitzer Survey of Stellar Structure in Galaxies} \citep{2010PASP..122.1397S}, gives 17\%, and SINGS, the {\it Spitzer Infrared Nearby Galaxies Survey} \citep{2005ApJ...633..857D,2007ApJ...655..863D}, giving 9\%. This third program was carried out during the cryogenic phase while the first two were conducted during the warm Spitzer mission. The new {\it Cosmic Flows with Spitzer} program has only contributed 3\% of the current data.  Smaller programs during the cryogenic mission supply us with information on the remaining galaxies.  The information comes from a multitude of programs but the integration times are the same within a factor two (mostly 240 sec, occasionally 120 sec).  The integrations are sufficiently deep to reach a surface brightness 26.5 mag/sq. arcsec (AB) even with the shorter exposures.  Details on programs and exposure times are included at the Extragalactic Distance Database website by selecting the catalog {\it Spitzer [3.6] Band Photometry}.\\

The photometric reductions were carried out by two independent procedures.  The method utilized by the CHP uses software developed for the GALEX Large Galaxy Atlas (Seibert et al. 2012, in prep.).  The method developed in anticipation of the arrival of CFS data is based on the Archangel photometry package \citep{2007astro.ph..3646S} described by \citet{2012AJ....144..133S} (hereafter SCT12) and earlier by \citet{2011MNRAS.415.1935C} in the context of optical photometry.  In a comparison of 171 galaxies (SCT12), the two procedures result in agreement at the level of 0.01 mag with rms scatter of 0.052.  Partitioned equally, the internal uncertainty (reductions of the same data by different methods and individuals) is $\pm0.037$ mag.  There are marginal differences for galaxies brighter than [3.6]=11 (CHP brighter at the level of 0.03) which is probably attributable to sky settings. We choose to average over CHP and CFS photometric values. \\

	Uncertainties on apparent magnitudes have been shown to be very small, cumulatively $\pm 0.05$ (SCT12).   The photometry reaches isophotal levels that require only a few percent extrapolation to give total magnitudes and the scale is stable to better than 0.01 mag across the sky (IRAC Instrument Handbook V2.0, 2011).  Setting the sky remains a dominant uncertainty at a level of 0.04 mag.  IRAC ch.1 [3.6] luminosities receive the following corrections: 
	
	1) $A_b^{[3.6]}$: galactic extinction \citep{1989ApJ...345..245C,1998ApJ...500..525S}, 
	
	2) $A_i^{[3.6]}$: internal extinction \citep{1995AJ....110.1059G,1997AJ....113...22G, 1998AJ....115.2264T}, 
	
	3) $A_k^{[3.6]}$: shift in the flux due to Doppler effect \citep{1968ApJ...154...21O, 2007ApJ...664..840H}, 
	
	4) $A_a^{[3.6]}$: extended emission from the Point Spread Function outer wings and from scattered diffuse emission across the IRAC focal plane  \citep{2005PASP..117..978R}. 
	
	These corrections are all discussed in SCT12.
The resulting apparent magnitude in the AB system is 
\begin{equation}
	 [3.6]^{b,i,k,a}= [3.6] - A_b^{[3.6]} - A_i^{[3.6]} - A_k^{[3.6]}+A_a^{[3.6]}.
\end{equation}

The data that are used in the following discussion are collected into Table~\ref{tbl:online}.  This table, the complete version given with the online publication, includes CFS and CHP total magnitudes, each including the 4 adjustments just described, and averages of the two methods.  The table also gives inclination and linewidth information drawn from TC12 and color terms for color corrections  described in sub-section 3.3.  The galaxies in Table~\ref{tbl:online} are either part of the zero point calibration (sample ZeroPt) or a member of a cluster contributing to the slope template. 


\section{[3.6] Band Calibration}

The TFR calibration requires the definition of a slope and the establishment of an absolute scale.
The slope is the trickiest item because there is a correlation between its value and a form of Malmquist bias.   Given two galaxies at the same distance with the same linewidth, the brighter galaxy might be chosen but not the fainter one.  The potential bias depends on the slope of the correlation because with a relatively flat slope most intrinsically luminous galaxies lie above the correlation while with a very steep slope these same galaxies tend to lie below the correlation.  Consider a target for a distance measurement in the field that intrinsically lies above the assumed mean relation, the trend for distant galaxies if the relation is flat.  With the distance measurement the target is assigned the mean luminosity of the correlation at the target's linewidth so given a distance that is too small. This bias has repeatedly been discussed at length, most recently by TC12.  The salient point is that the so-called `inverse' relation (ITFR), the least squares regression where errors are taken to be in linewidth only, gives results that are close to bias free.  \citet{1994ApJS...92....1W} pointed out that, while in his experiments the ITFR bias was reduced by a factor 6 from that incurred using the direct relation, yet a small bias remained because the sample selection was not made in the band he considered. We have the same problem.  Our strategy is to use the ITFR and then evaluate the bias with simulations anticipating that, like with the $I$ band calibration, the effects will be small.  The bias tests are discussed in a later section.\\

	The calibration process has been described in detail by \citet{2000ApJ...533..744T}  and TC12. With the $I$ band relation there is no clear evidence for scatter due to a third parameter but the situation at [3.6] is different.  A color term is found and that matter will be discussed.

\subsection{Relative distances and ITFR Slope}

\begin{figure}[h!]
\includegraphics[scale=0.45]{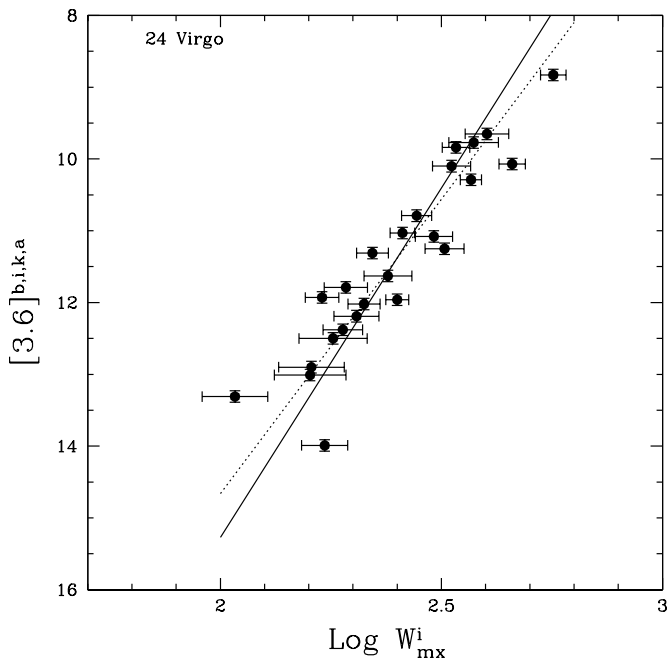}
\caption{Tully-Fisher relation in the [3.6] band for the Virgo Cluster. The solid line gives the inverse fit of the universal template correlation. The dotted line is the inverse fit of the correlation for the Virgo Cluster alone.}
\label{Vir}
\end{figure}

\begin{figure}[h!]
\includegraphics[scale=0.7]{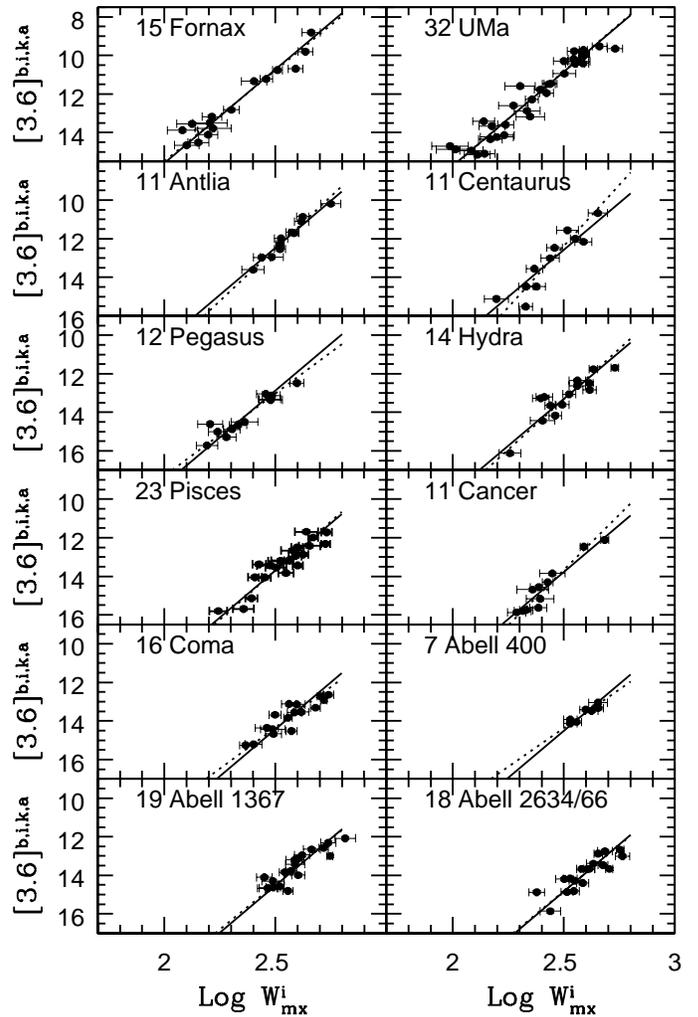}
\caption{Tully-Fisher relation in the [3.6] band for the Fornax, Ursa Major, Antlia, Centaurus, Pegasus, Hydra, Pisces, Cancer, Coma, Abell 400, Abell 1367 and Abell 2634/66 clusters. Solid lines give the inverse fit of the universal template correlation. Dotted lines are the inverse fits of the correlation for each cluster.}
\label{6clusters}
\end{figure}

	The measurement of distances requires the hypothesis of a universal correlation. To begin, we make inverse fits to each one of the clusters separately. Dotted lines in Figures \ref{Vir} and \ref{6clusters} illustrate the inverse fits of the TFR for each cluster. Slopes are quite similar between clusters.  Slopes and their uncertainties are given for each cluster in Table \ref{tbl:clfits}.  The individual fits are consistent with the soon to be derived best fit and hence with the universal correlation hypothesis.  As cluster distances increase, the faint luminosity limits increase. However, no dependence of the slope with distance is seen, as would be a marker of Malmquist bias (we still make a tiny correction for bias to cluster moduli as described in Section 3.4).\\

The next step is to combine the 13 individual cluster correlations by vertical translations.  The Virgo Cluster is used as a reference.  Each preliminary zero point from the individual fits provides us with a first estimate of the relative distance between the Virgo Cluster and the cluster in question.\\ 

\begin{figure}[h!]
\includegraphics[scale=0.45]{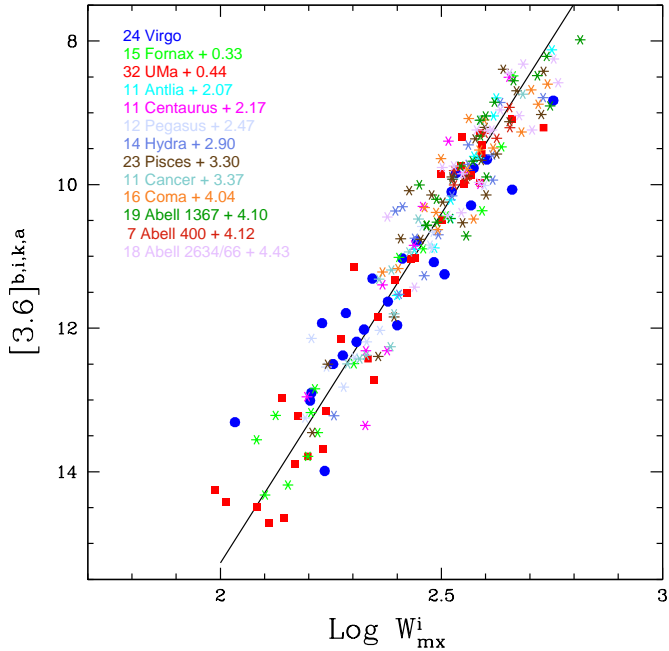}
\caption{Template Tully-Fisher relation in the [3.6] band obtained with data from 213 galaxies in 13 clusters. Offsets given with respect to the Virgo Cluster represent distance modulus differences between each cluster and Virgo. The solid line is a least squares fit to all the galaxies with errors entirely in linewidths, the ITFR.}
\label{TF}
\end{figure}

  	Apparent magnitude zero points confirm that Virgo, Fornax and Ursa Major are the closest clusters. Then come Antlia-Centaurus-Pegasus, then Hydra-Pisces-Cancer, and finally Coma and the three Abell clusters A1367, A400 and A2634/66. To establish the best universal slope and the best relative distances between clusters, we follow an iterative procedure.  We initially consider the nearest three clusters because they are observed to comparable depths in intrinsic magnitude.  The Fornax and Ursa Major magnitude scales are shifted according to the difference in zero point with respect to Virgo.  A least squares fit of the ITFR is made to this ensemble.  The new slope is assumed in a fit to the 3 individual clusters with only the zero point as a free parameter in each case. Given the new zero point offsets the cycle is repeated, leading to rapid convergence.  This procedure is repeated with the addition of each distance group in turn.  Again, convergence is rapid.  It is to be stressed that this procedure works because, following expectations, the slope of the ITFR is not affected by the magnitude level of truncation.  This procedure would manifestly {\it not} work with the direct or bi-variate relations where the slopes vary with the level of truncation.  In the end we obtain a slope of $-9.74\pm0.22$ for the template ITFR. Zero point offsets with this `universal' slope are shown in Figure \ref{TF} and give relative distance moduli of clusters referenced to the Virgo Cluster. The universal slope of the ITFR is displayed in Figure \ref{TF}  as well as by the solid lines in Figures \ref{Vir} $-$ \ref{6clusters}.

\subsection{Zero Point and Absolute Distances}

\begin{figure}[h!]
\includegraphics[scale=0.45]{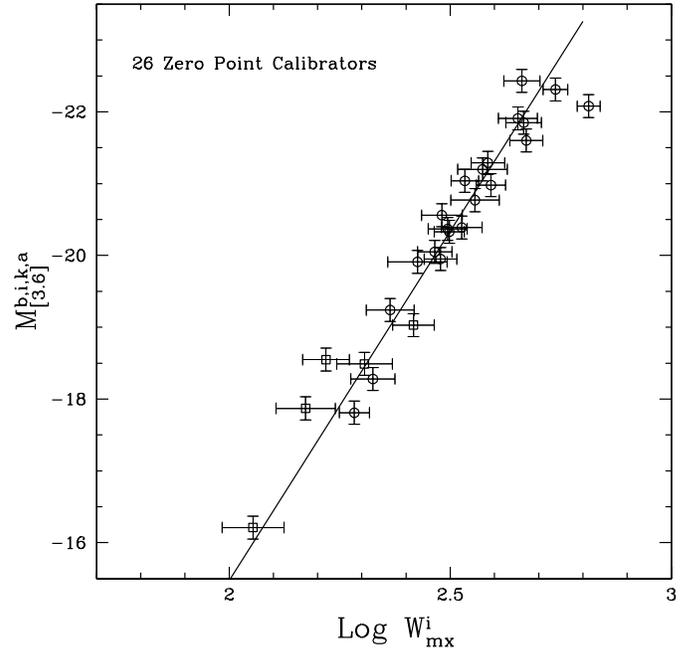}
\caption{TFR for the 26 galaxies with distances established by observations of Cepheid stars (circles) or the Tip of the Red Giant Branch (squares). The solid line is the least squares fit with the slope established by the 13 cluster template. The zero point of the TFR is set at the value of this fit at ${\rm log} W^i_{mx} =2.5$.}
\label{ZP}
\end{figure}

Presently, [3.6] photometry is available for 26 nearby galaxies with suitable morphologies, inclinations, and linewidths that also have well measured distances from either the Cepheid period-luminosity or tip of the red giant branch methodologies.  These 26 are a subset of the 36 absolute calibrator galaxies used in the $I$ band calibration (TC12).   Their luminosity-linewidth correlation is seen in Figure~\ref{ZP} where now the ordinate is absolute magnitudes from the established distances.  The line is a least squares fit with the slope  $-9.74$ prescribed by the template.  The zero point is $- 20.34\pm0.10$.  The most deviant point is the fastest rotator, NGC 2841, with a deviation of $2.7 \sigma$ with respect to the template dispersion.  This galaxy was a $2.3 \sigma$ deviant in the $I$ band calibration.  There is nothing unusual about this galaxy other than its extreme rotation rate so we see no reason to disregard it as a calibrator.\\

\begin{figure}[h!]
\centering
\includegraphics[scale=0.45]{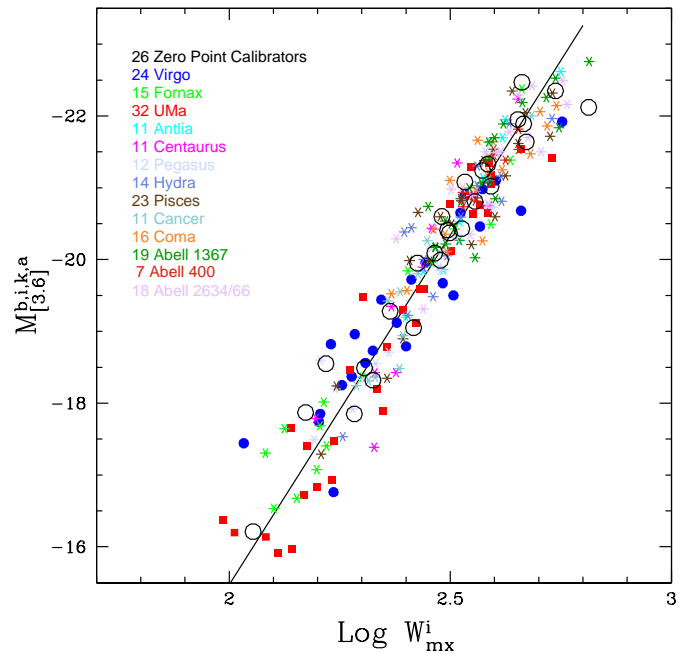}
\caption{The template of the new [3.6] band - HI linewidth correlation is built with 213 galaxies in 13 clusters extending in range from 1000 to 10,000 km s$^{-1}$ with the absolute magnitude scale set by 26 zero point calibrators.}
\label{TFtot}
\end{figure}

The distance to the Virgo Cluster is given by the zero point of the constrained slope shown in Figure~\ref{TF} minus the zero point of the absolute calibration shown in Figure~\ref{ZP}.  Application of this shift allows both cluster template and zero point calibrator galaxies to be plotted together as seen in Figure~\ref{TFtot}.
The ITFR expression in the [3.6]-band is given by:
\begin{equation}
M^{b,i,k,a}_{[3.6]} = -( 20.34\pm0.10) -( 9.74\pm0.22) ({\rm log} W_{mx}^{i} -2.5)
\end{equation}
\\

The TFR scatter in magnitudes (relevant for distance measurements) is given by:
\begin{equation}
\sigma_{TF}=\sqrt{\frac{\chi^2}{N-1}}
\end{equation}
where  $\chi^2$ is the minimum of 
$$\sum(M_i- (a + b({\rm log} W_{mx}^{i} -2.5) ))^2$$ 
with $a$ and $b$ the zero point and slope of the ITFR respectively and $N-1$ are the degrees of freedom.
The scatter for the entire cluster template sample is $\pm0.49$ mag from the universal ITFR, corresponding to a scatter in distance of 25\%. The scatter for the 26 zero point calibrators is a similar 0.44 mag.  Dispersion increases toward fainter magnitudes as well documented at $I$ band by \citet{1997AJ....113...53G}.  The sample presented here is still limited but the dispersion is consistent with a Gaussian distribution.  With large samples \citep{2008ApJ...676..184T} one finds about $3\%$ of candidates are more deviant than anticipated by Gaussian statistics.  The causes are not always evident.\\

Scatter may arise from: 1) measurement uncertainties affecting magnitudes, inclinations, and linewidths, 2) correction uncertainties applied to measured parameters, and 3) `cosmic' scatter, e.g. cluster depth effects or interlopers, deviations from disk planarity, other gravitational and photometric asymmetries, variations in the stellar population make-up, variations in disk-to-bulge ratios, etc.  Whatever the sources, we have a standard to meet set by the $I$ band analysis.  The samples used in the current analysis involve 80\% of the samples used in the $I$-band calibration (TC12).  Inclinations and linewidths are the same, the factors mentioned associated with cosmic scatter are the same, corrections to photometric parameters are reduced in the mid-infrared, and the integrity of the magnitude measurements must be at least as good or better with the Spitzer observations since observations are made all-sky with the same instrumental configuration.   Error bars on magnitudes are reduced in  Figures $1-5$ compared with those on equivalent plots in the $I$ band calibration paper (TC12) to the degree that observational errors in magnitudes are a minimal component of uncertainties. Yet the scatter found at $I$ band is less: $\pm0.41$ mag for the cluster template sample, lower with a significance of $2 \sigma$,  and 0.36 mag for the zero point calibrators.   As much as half of the increase in magnitude scatter will occur because the slope of the correlation is steeper in the mid-infrared.    However there could be an additional explanation for the increased scatter found at [3.6].

\subsection{A Color Term}

\begin{figure}[h!]
\includegraphics[scale=0.55]{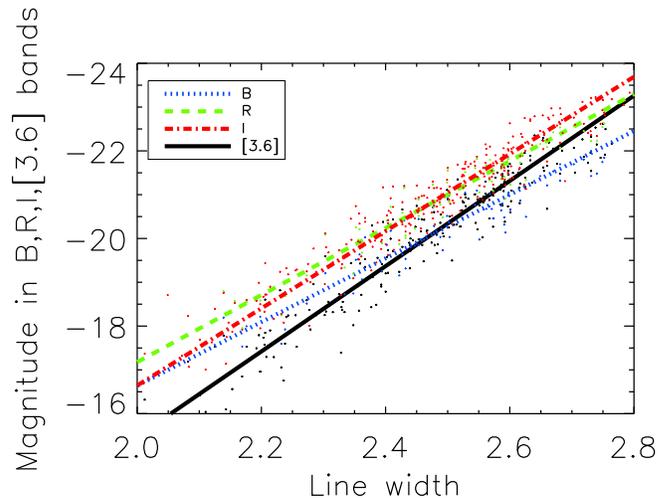}
\caption{TFR in $B, R, I$ and [3.6] bands. $B$ and $R$ bands data are from \citet{2000ApJ...533..744T} , $I$ band data are from  TC12 and [3.6] band data are from SCT12. Linewidths are the same as used by TC12. The slopes steepen from blue to red, with values $-7.27$ at $B$, $-7.65$ at $R$, $-8.81$ at $I$, and $-9.74$ at [3.6].}
\label{slope}
\end{figure}

\begin{figure}[h!]
\centering
\includegraphics[scale=0.456]{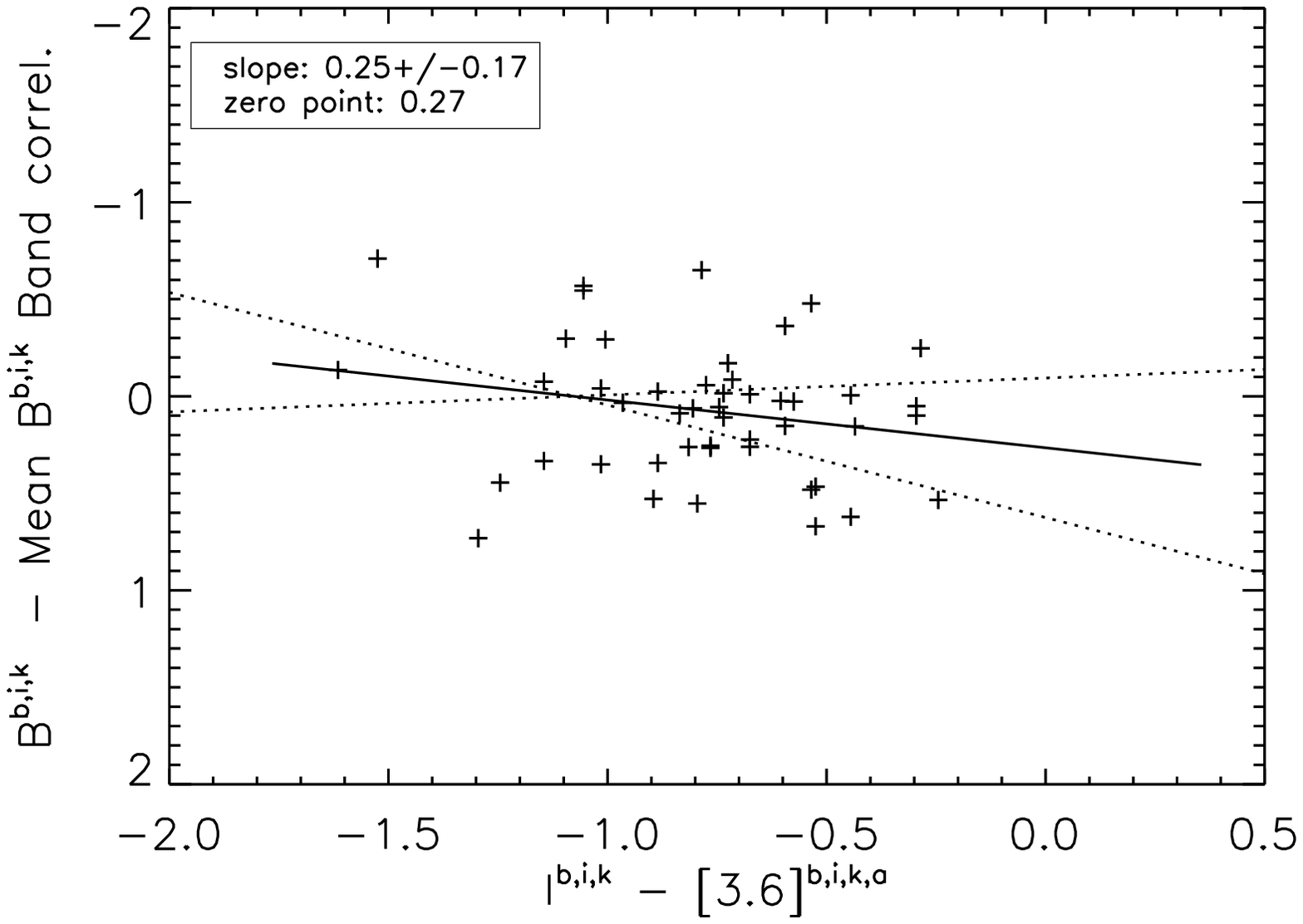}
\includegraphics[scale=0.456]{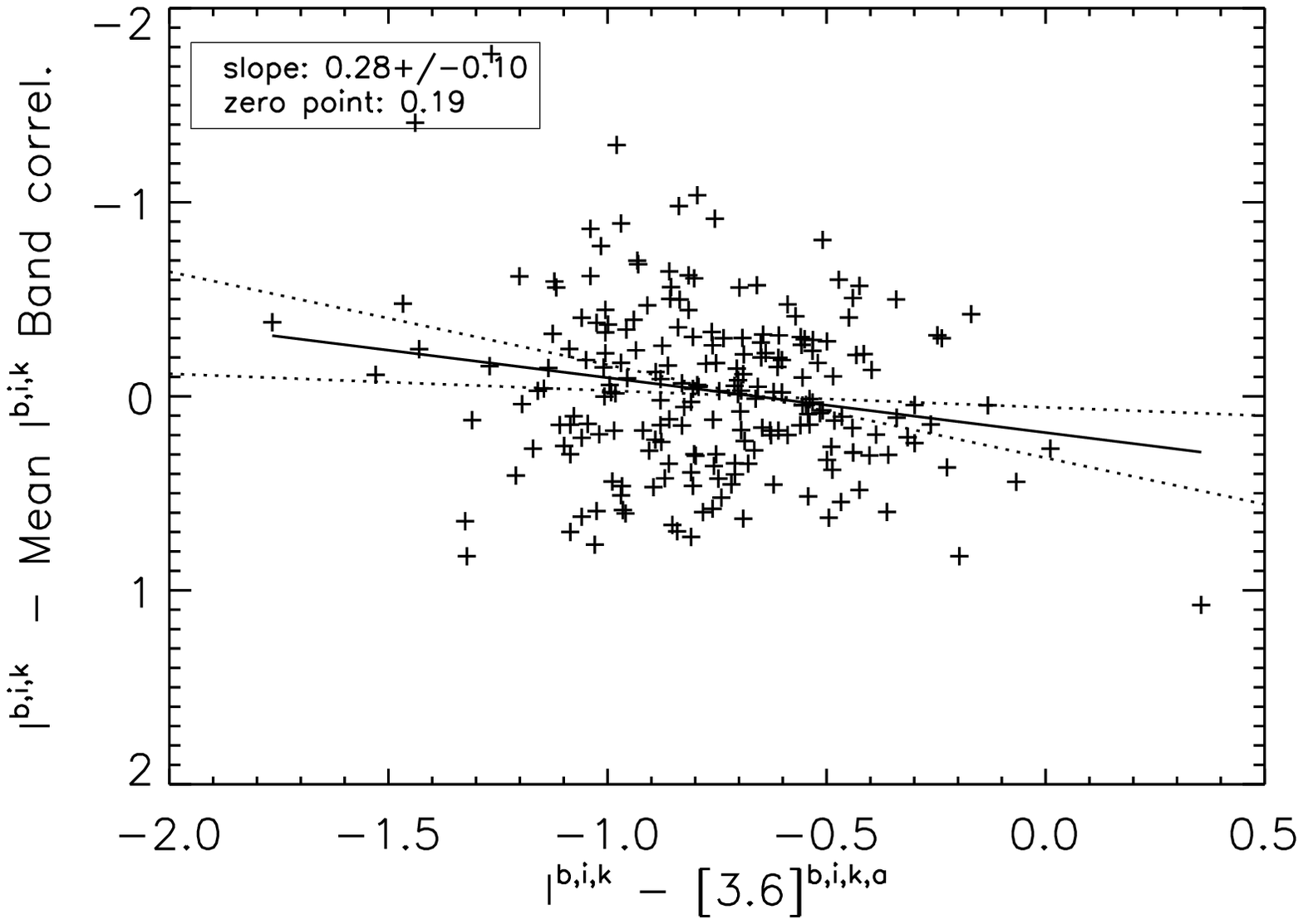}
\includegraphics[scale=0.456]{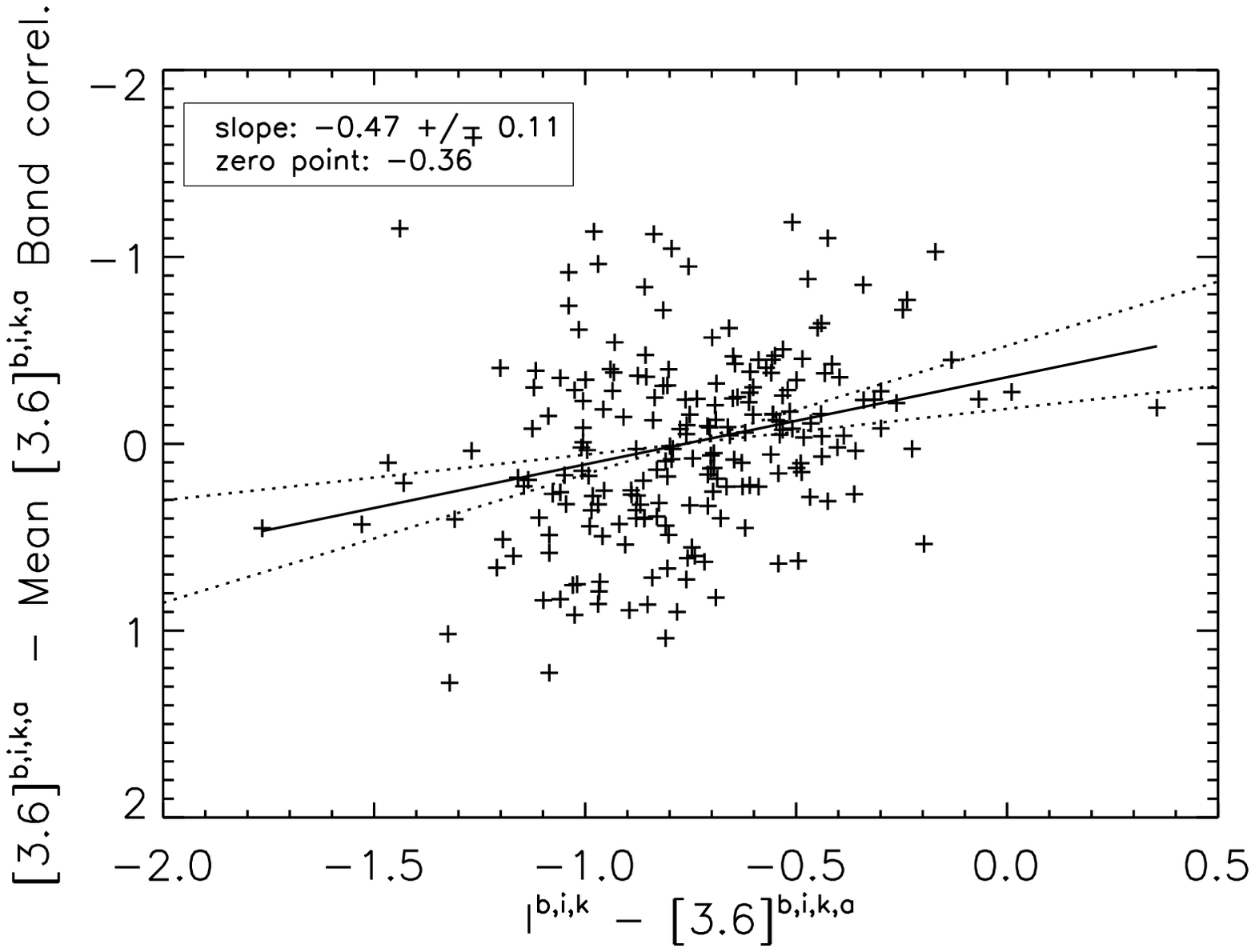}
\includegraphics[scale=0.456]{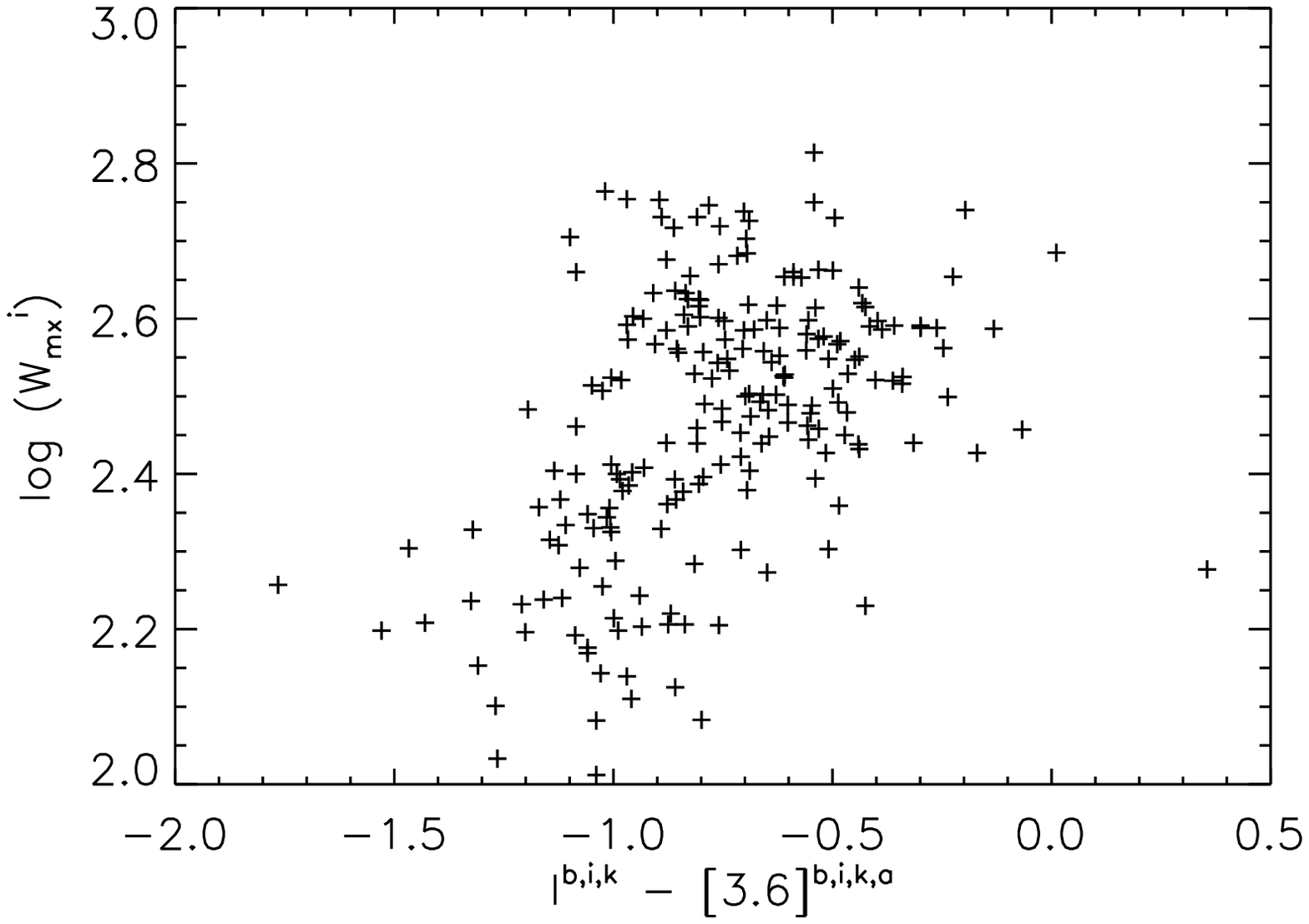}
\caption{{\it Top 3 panels:} Deviations from the mean ITFR relation as a function of $I-[3.6]$ color. Solid and dotted lines are best fits and 95\% probability limits.  {\it Top:} At $B$ band red galaxies tend to lie below the mean relationship. {\it Top middle:} At $I$ band there is a hint that red galaxies lie low although the correlation fit is dominated by a few extreme cases.  {\it Bottom middle:} At [3.6] band the sense of the correlation has flipped and red galaxies tend to lie above the mean relation. {\it Bottom:} The correlation between linewidth and color.}
\label{color}
\end{figure}

\begin{figure}[h!]
\centering
\includegraphics[scale=0.53]{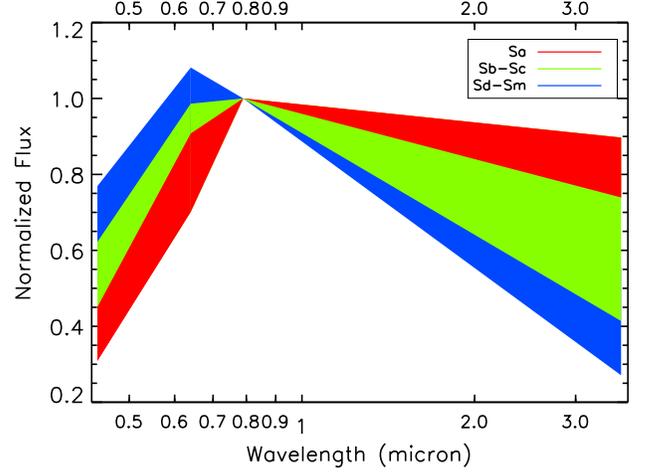}
\caption{Representation of fluxes at $B,R,I$,[3.6] bands normalized to unity at $I$ band.  Type Sa: red; types Sb-Sc: green; types Sd-Sm: blue.  The extrema are defined by members of our sample and color swaths indicate the domains dominated by the different types.}
\label{sed4band}
\end{figure}

It has long been known that the TFR steepens toward longer wavelengths \citep{1982ApJ...257..527T}.   The effect is seen in Figure~\ref{slope}. (Note: in the discussions in this section all optical photometry values have been transferred from Vega to AB zero points.) There is  a strong color correlation with linewidth, more rapidly rotating galaxies tend to be redder, so at longer wavelengths the high rotation end of the TFR rises with respect to the low rotation end.  Within a small linewidth interval, redder galaxies will rise more than bluer galaxies.  It follows that red and blue galaxies cannot be well mixed in the TFR at all wavelengths.   The trends that could be anticipated are shown in Figure~\ref{color} (only a portion of the sample have photometric measurements at $B$ band).  The comparison of fluxes at four bands from $B$ to [3.6] for individual sources given in Figure~\ref{sed4band} confirms the well known linkage between galaxy type and color.  Early type galaxies have relatively more infrared flux relative to late type galaxies.  This point was also illustrated with the representative spectral energy distribution plots in Figure~1 of SCT12.  Galaxies that are more luminous and earlier in type are dominated by older, more metal enriched red giant stars emitting more in the infrared.

\begin{figure}[h!]
\centering
\includegraphics[scale=0.45]{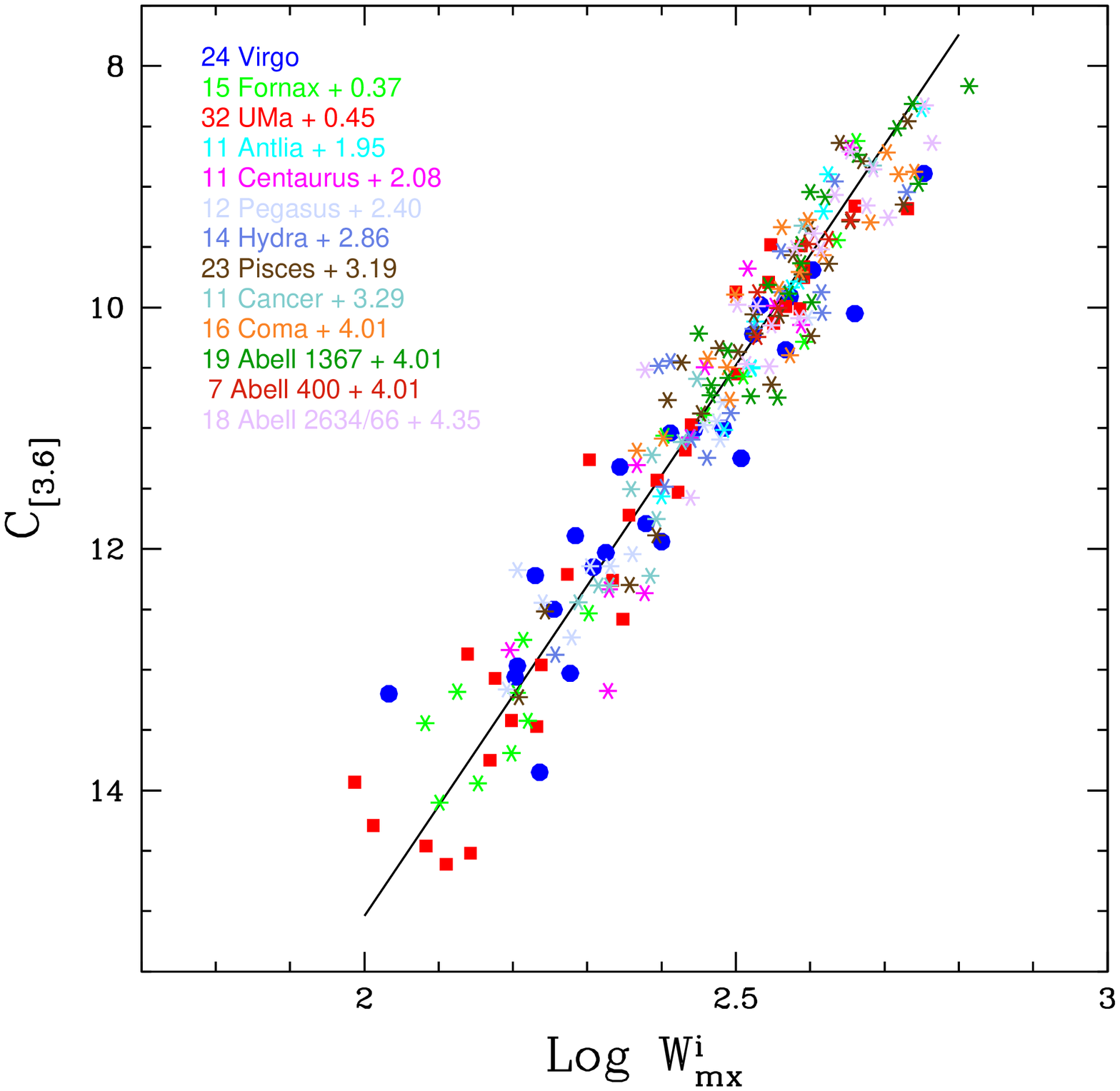}
\includegraphics[scale=0.45]{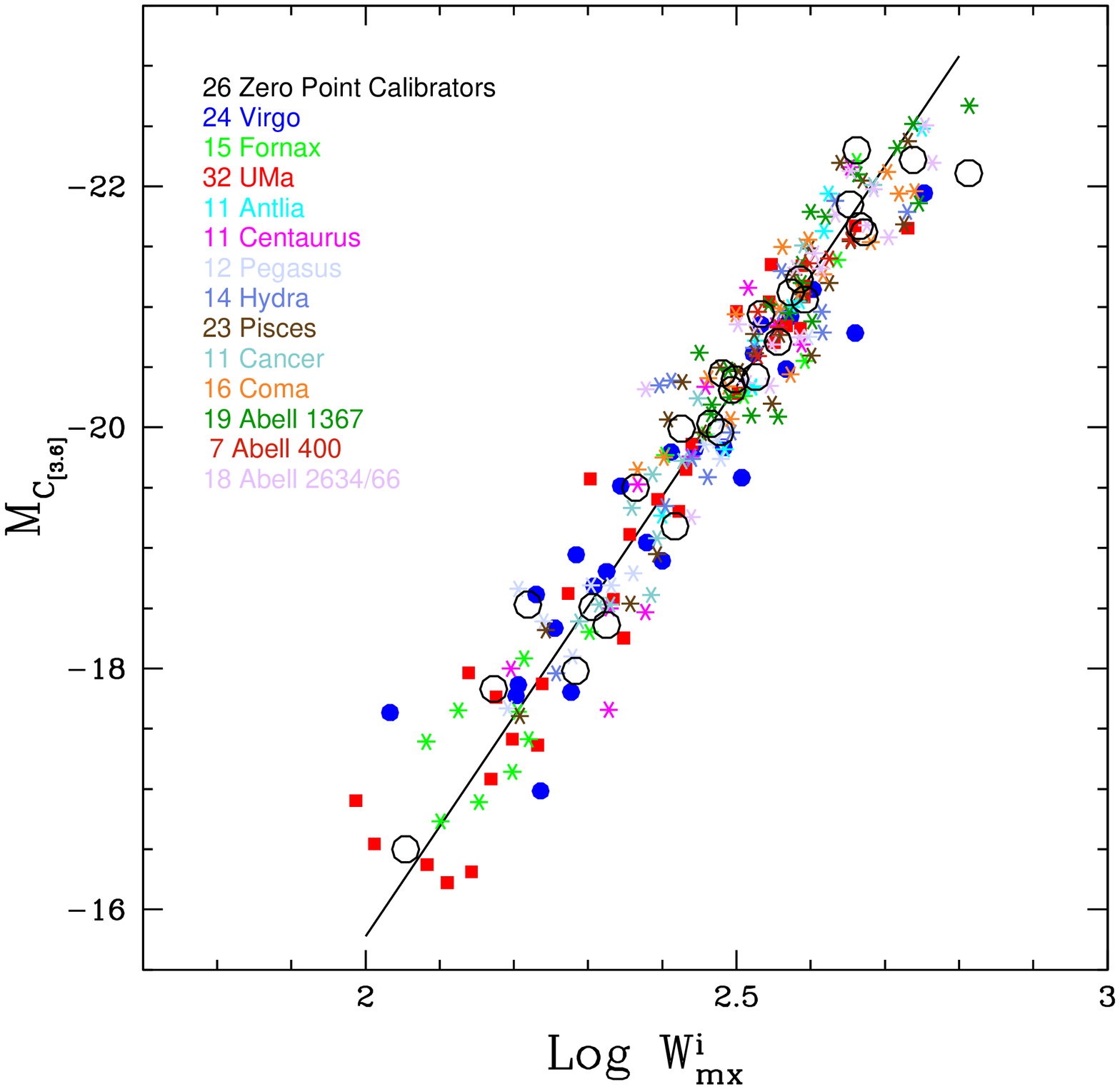}
\caption{The ITFR after adjustments for the color term. {\it Top:} Color adjusted apparent magnitudes translated to the relative distance of the Virgo Cluster.  {\it Bottom:} Color adjusted absolute magnitudes with the absolute distance scale established by the galaxies with independent distances represented by large open circles.}
\label{TFLambda}
\end{figure}

There have been long standing suggestions that the dispersion in the TFR might be reduced by inclusion of additional parameters.  In an early instance \citep{1985ApJ...289...81R}, when only photographic or photoelectric magnitudes were available,  the case was framed in terms of galaxy types which are strongly correlated with color.  \citet{2006ApJ...653..861M} have maintained the use of a type separation with $I$ band work.  \citet{2000ApJ...533..744T} acknowledged the hint of a type dependence in the $I$ band relation but concluded that the evidence remained too weak to warrant adding complexity to the TFR analysis.

The situation changes with the mid-infrared information.  In spite of superior photometry the scatter in the TFR is increased and there is a significant color signature.  The variations in spectral energy distribution implicit in the range of representative colors shown in Figure \ref{sed4band} provide a natural explanation given the extended lever arm from the optical to the [3.6] band.\\

There is also the possibility that some flux in the [3.6] band may come from other than old stars.  \citet{2012ApJ...744...17M} determined that $12\pm5\%$ of [3.6] flux arises from hot dust, PAH emission, or young to intermediate age stars in 6 representative spiral galaxies observed with {\it Spitzer Space Telescope}.   However the variance of 0.05 mag is small compared with the ITFR scatter.  Moreover, it  can be anticipated that the galaxies most affected by manifestations of star formation are later, bluer types, whence augmented flux will tend to diminish a color term arising from old stars.\\

Whatever the cause, it can be anticipated that the scatter can be decreased with the introduction of a color correction.  To address this issue we consider the straight line fits included in the top three panels of Figure~\ref{color}.   The fits are least squares minimizations on the ordinate parameter; the difference in magnitude of a target from the mean TFR.  The bottom panel shows the concordant variation of color with linewidth.  Faster rotators tend to be redder.\\

In the mid-infrared case, the offset for an individual galaxy from the mean fit in the figure  is:
\begin{equation}
\Delta M_{[3.6]}^{color} = M_{[3.6]}^{b,i,k,a} + 20.34 + 9.74 ({\rm log} W^i_{mx} - 2.5).
\label{delm}
\end{equation}
An equivalent correction can be constructed with apparent magnitudes rather than absolute magnitudes, $\Delta [3.6]^{color} = \Delta M_{[3.6]}^{color}$, with an appropriate replacement of the zero point constant in Eq.~\ref{delm}.
The correction term commensurate with the fit in the third panel of Figure~\ref{color} is
\begin{equation}
\Delta [3.6]^{color} = \Delta M_{[3.6]}^{color} = - (0.47\pm0.11) [(I-[3.6]) + 0.77].
\end{equation}
We introduce a new color adjusted magnitude parameter $C_{[3.6]} = [3.6]^{b,i,k,a} - \Delta [3.6]^{color}$ where the distinct nomenclature emphasizes the composite nature of this pseudo-magnitude.  Next, the analysis discussed in Section 3.2 leading to the construction of Figure 4 is repeated.  Likewise, the adjustments are applied to the calibrators with independently established distances and the procedures are repeated that lead to Figure~\ref{TFtot}.  The adjusted relations are shown in Figure~\ref{TFLambda}.  The new correlation is described by the formula:
\begin{equation}
M_{C_{[3.6]}} = -(20.34\pm0.08) -( 9.13\pm0.22) ({\rm log} W_{mx}^{i} -2.5)
\end{equation}
The flattening of the adjusted relation comes about since redder systems move downward and redder galaxies tend to have larger linewidths.
The overall magnitude scatter in the new relation is $\pm0.44$ mag (corresponding to a scatter in distance of 22\%), down from 0.49 mag before adjustment, and comparable with 0.41 found at $I$ band with an otherwise comparable analysis (TC12).  The comparable numbers for the zero point calibrators alone are a scatter of 0.37 with the adjusted parameter $C_{[3.6]}$,  0.44 before the adjustment, and 0.36 at $I$ band.  The comparisons between [3.6] and $I$ have some imprecision because the sample sizes for the latter are 25\% greater.  The TFR parameters derived from alternative samples and bandpasses are summarized in Table~\ref{tbl:compare}.

\subsection{Bias}

\citet{1994ApJS...92....1W} showed that a small Malmquist bias exists in the use of the ITFR, although reduced from the direct TFR by  a factor of 6 in the situation he explored \citep{1995ApJ...446...12W}, reducing the bias reflected in the Hubble Constant from $17\%$ to $3\%$.  The bias arises from two effects.  First, sample selection departs from an idealized case of a flat magnitude limit because samples have been selected in blue bands and color terms translate to a slope in the limiting magnitude in the infrared: slower rotators which tend to be bluer are favored for inclusion over faster rotators which tend to be redder (see Figure \ref{color}, bottom).  Second, the shape of the galaxy luminosity function contributes to the bias because there are more intrinsically fainter galaxies that scatter bright-ward through errors than intrinsically brighter galaxies that scatter faint-ward \citep{1913MNRAS..73..359E}.  The bias increases with distance as the effect of the exponential cutoff of the luminosity function plays an increasing role.\\

\begin{figure}[h!]
\centering
\includegraphics[scale=0.45]{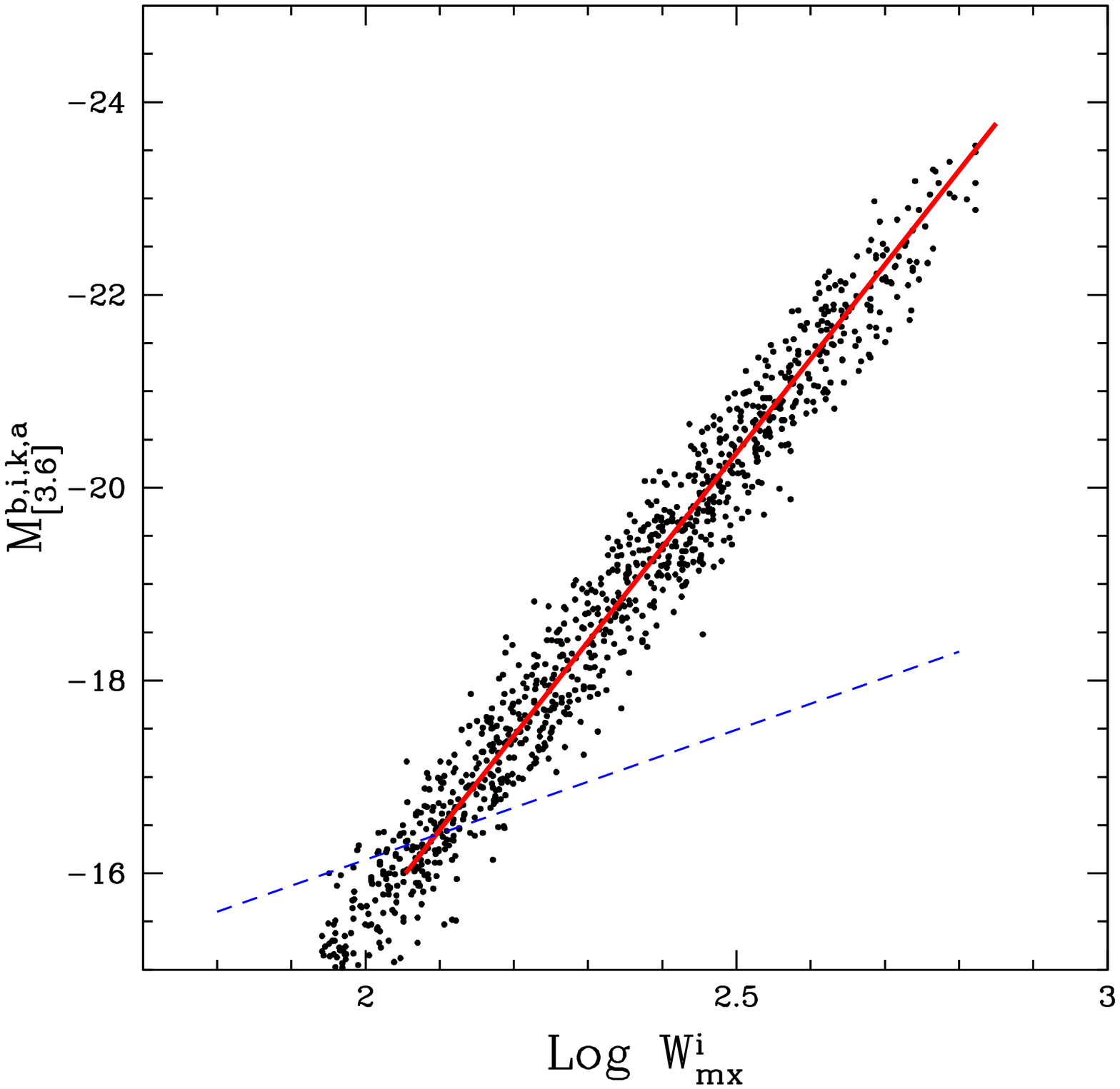}
\includegraphics[scale=0.45]{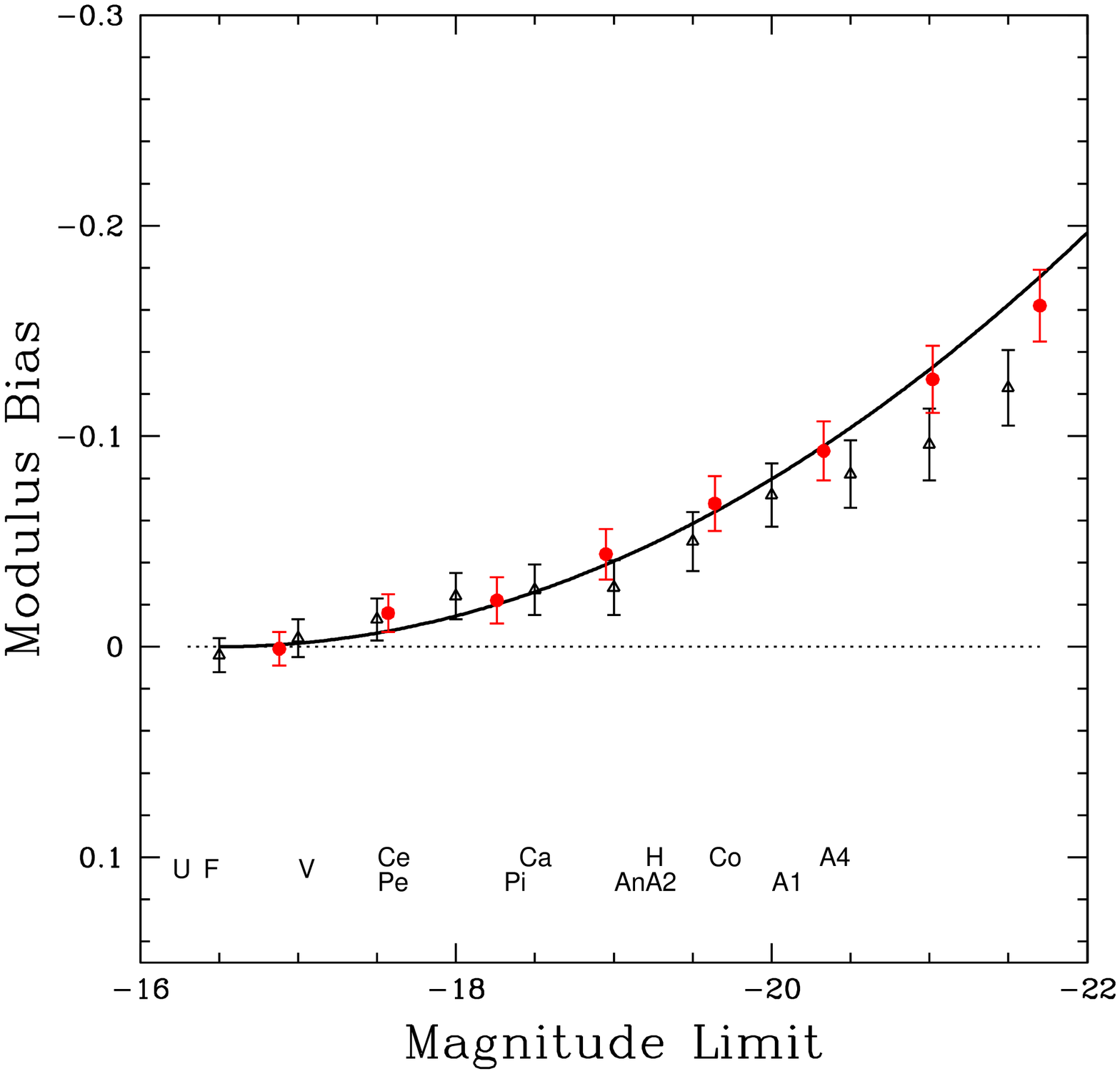}
\caption{{\it Top:} Simulated TFR drawing randomly from a Schechter luminosity function with slope $\alpha = -0.9$ and cutoff $M^{\star} = -22$.  The ITFR has slope $-9.13$ and scatter 0.4 mag.  The dashed blue slanting line illustrates the color dependence at the faint limit resulting from sample selection in the blue.  {\it Bottom:} Bias $<\Delta M>_{measured}$ as a function of absolute magnitude limit which increases with distance.  Black triangles: flat faint limit; red circles: faint limit increasing with increasing linewidth in accordance with blue line in top panel.  Solid curve: the empirical bias fit $b = -0.0065 (\mu -31)^2$. Letters at the bottom: codes for the 13 calibrating cluster (see Table~\ref{tbl:compIcal} for translation of codes). Their horizontal positions indicate sample limits and vertical intercepts with the solid curve give the corresponding biases.}
\label{bias}
\end{figure}

The amplitude of the bias from the two effects was explored with the calibration at $I$ band (TC12).  The situation now with the [3.6] band sample is slightly worse than at $I$ because the wavelength interval from selection at $B$ is larger.  The bias analysis carried out in the case of the $I$ band calibration is repeated here, tailored to the current situation.  We first combine the Virgo, Fornax, and Ursa Major samples to improve statistics and include contributions from a range of environments.  This ensemble is described by a \citet{1976ApJ...203..297S} function with faint end slope $\alpha = -0.9$ and a bright end cutoff at $M_{[3.6]}^{\star} = -22$.  Then we randomly populate an artificial TFR to match the observed [3.6] band relation, drawing from the Schechter luminosity function.  The faint limit is determined empirically to roughly obey the relation $M_{[3.6]}^{lim} = C_{\ell} - 2.70 ({\rm log} W^i_{mx} -1.8)$ where $C_{\ell}$ couples with distance.  The artificial TFR and the cutoff for the nearest clusters is shown in the top panel of Figure~\ref{bias}.  The dashed blue line indicates the cutoff experienced at a distance modulus of 31.  The cutoff, characterized by $C_{\ell}$, slides to brighter (more negative) magnitudes linearly with increasing distance modulus.  The bias $<\Delta M>_{measured}$  is determined at intervals of $C_{\ell}$ corresponding to increasing distance.  Here, $<\Delta M>_{measured}$ is the average deviation from the fiducial relation where $<\Delta M>_{true} = 0$ by construction.  The growth of the bias as a function of cutoff magnitude is seen in the bottom panel of Figure~\ref{bias}.  The solid curve, normalized to unity at a distance modulus $\mu = 31$ where even the faintest of useful candidates are included, is described by the formula between bias, $b$, and distance modulus, $\mu$:
\begin{equation}
b = -0.0065 (\mu - 31)^2 .
\end{equation}
By comparison, the coefficient in the case of the $I$ band analysis is $-0.005$.  The letters at the bottom of the figure are codes for the 13 calibrating clusters (see Table~\ref{tbl:compIcal} to decipher codes) and their horizontal placements indicate the respective sample limits and projection upward gives the corresponding biases.  These biases are recorded in Table~\ref{tbl:clfits} and are reflected in the adjusted cluster moduli and distances.  For a galaxy in the field, the corrected distance modulus $\mu^c$ can be expressed as 
\begin{equation}
\mu^c = (C_{[3.6]} - M_{C_{[3.6]}}) + 0.0065[(C_{[3.6]} - M_{C_{[3.6]}}) - 31]^2  
\end{equation}

\section{The Hubble Constant}

\begin{figure}[h!]
\includegraphics[scale=0.55]{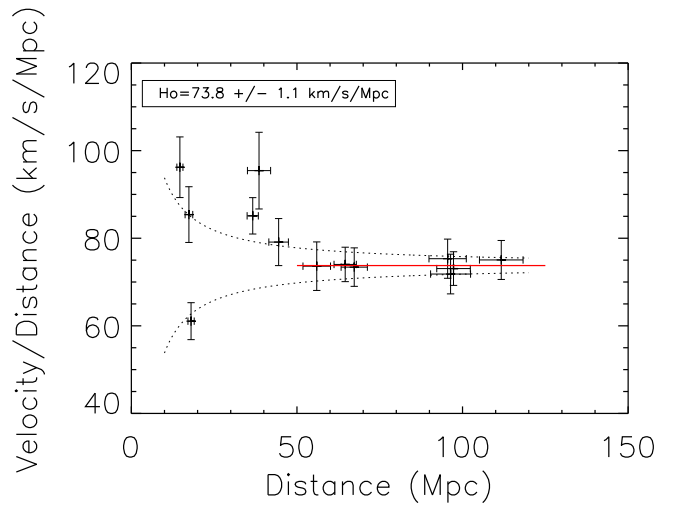}
\caption{Hubble parameter as a function of distance. The solid line is a fit to cluster points at distances greater than 50 Mpc ($V_{CMB} > 4000$ km s$^{-1}$). The fit gives H$_0 = 73.8 \pm 1.1 $ km s$^{-1}$ Mpc$^{-1}$.  Curved dotted lines illustrate deviations in velocity of 200~\kms\ from the fit.}
\label{hp}
\end{figure}

The last column in Table~\ref{tbl:clfits} records the `Hubble parameter' for each cluster: the velocity of the cluster in the CMB frame divided by the measured distance.  These quantities are plotted against distance in Figure~\ref{hp}.  A similar figure was presented as a summary of results  from the $I$ band calibration with the same 13 clusters (TC12: distances compared in Table~\ref{tbl:compIcal}).  Here, as there, we see a large scatter in the Hubble parameter for the nearer clusters and small scatter for the more distant clusters.  It can be anticipated that the measures for the nearer clusters are strongly affected by peculiar motions.  The 5 clusters within 40~Mpc are all part of our extended supercluster complex: either within the historic Local Supercluster or the so-called Great Attractor region.  The low scatter among the 7 clusters more distant than 50 Mpc ($V_{CMB} > 4000$~\kms) suggests that the relative contributions of peculiar velocities have a modest effect on redshifts at such large distances.\\

In the case of the $I$ band calibration, the mean value of the Hubble parameter for the 7 most distant clusters was $75.1 \pm2.7$ \kmsMpc\ where the error is just the rms scatter of the 7 contributions.   That value would increase to 75.8 with the revised LMC distance from Monson et al (2012).  With the present calibration, including the new LMC distance, the fit shown in Figure~\ref{hp} gives a value of H$_0 =73.8$ with an rms scatter of 1.1 and a standard deviation of 0.4~\kmsMpc\ for the same 7 clusters considered previously.  If the fit is extended to include the Pegasus Cluster at 44.5~Mpc then H$_0 = 74.4$ and the scatter is 2.0~\kmsMpc.  The effect of a deviant radial motion of 200~\kms\ is illustrated in the figure as a function of distance.\\  

The uncertainty from the fit in Figure~\ref{hp} is given by the statistics of the deviations of the 7 contributions and is unrealistically low.  This error is what is expected if there is perfect Hubble expansion.  If peculiar motions of 200~\kms\ are the norm, and given the expected statistical errors in the distance of each cluster, the anticipated scatter around the mean Hubble value is $\pm2.6$~\kmsMpc.   We consider this to be our $1 \sigma$ random error.  We have several sources of systematic error.  The dominant component, creating almost 4\% uncertainty in H$_0$, comes from the uncertainty in the TFR zero point with just 26 calibrators.  Combined with a small uncertainty from the finite population of the template, the uncertainty in H$_0$ associated with the TFR calibration (assuming the zero point calibrator distances are perfect) is $\pm2.9$~\kmsMpc.  The zero point calibrator distances are not perfect. \citet{freedman2012} and \citet{2011ApJ...730..119R} report that with new Milky Way parallaxes for Cepheid stars and mid-infrared Spitzer photometry the uncertainty in the Cepheid scale is in the range $\pm [1.9-2.5]$~\kmsMpc.  The TRGB zero point calibration which  concerns 4 of the 26 calibrators, has similar or smaller systematics. The cumulative systematic error in H$_0$ is $\pm[3.5-3.8]$~\kmsMpc.  Combining random and systematic components we find H$_0 = 73.8 \pm2.6$(ran)$\pm[3.5-3.8]$(sys)~\kmsMpc. 


\section{Conclusions}

A great concern with studies of motions on large scales with the TFR has been the possibility that systematic errors in photometry could create spurious flows.  Small offsets between different observers, instruments, conditions, hemispheres, or seasons could be sky-sector dependent. Probably the single most important advantage of the use of space-based photometry such as offered by the Spitzer mission comes from the confidence that measurements are on the same scale at better than 1\% in all parts of the sky.
There are other advantages.  Obscuration is minimal both within targets and from our Galaxy.  This latter point is especially significant because studies of galaxy flow patterns can now reach high levels of completion across the sky.  Then it is a considerable advantage that the great majority of flux at [3.6] band arises from old stars, mainly those on the red giant branch. It can be surmised from the modest scatter in the TFR that there is a close coupling between the mass in stars and the dynamical mass.  And there is an advantage, at least vis \`a vis ground infrared observations, with the sensitivity achieved because of very low sky noise.  All but a few percent of the total flux is measured within isophots resolved from the  noise.\\

A small disadvantage with the mid-IR TFR calibration has been revealed with the documentation of a color term.  This color term is understood as the natural consequence of the correlation between galaxy rotation rate or luminosity and color \citep{1982ApJ...257..527T}.  At a given linewidth, red galaxies progressively get brighter relative to blue galaxies as one considers the TFR at longer wavelengths.  Evidence is accumulating that intrinsic scatter in the simple two parameter TFR is minimal with photometry at about $1 \mu$m.  A consequence of the color dependence is a steepening of the TFR toward the infrared.  If one is interested in the physical implications of the TFR rather than its use as a distance tool then the bivariate fit is of interest. Our template sample has the bivariate dependence $M_{[3.6]} \propto W^{3.8\pm0.1}$ which is 0.4 steeper than was found with almost the same sample at $I$ band (TC12).\\ 

At the expense of the requirement of extra knowledge in the form of a color, the TFR in the [3.6] band can be reformulated in a form with scatter that matches the best optical formulations.  The correction is small and not acutely dependent on the color measurement.  The appropriate ITFR equation for the measurement of distances is
$$M_{C_{[3.6]}} = -(20.34\pm0.08) - (9.13\pm0.22) ({\rm log} W_{mx}^{i} -2.5)$$
where $M_{C_{[3.6]}}$ is derived from the corrected apparent magnitude $[3.6]^{b,i,k,a}$ of a source minus the color term
$$\Delta [3.6]^{color} = -0.36 - 0.47 (I-[3.6]).$$
The slope of this formulation has been derived from a sample of 213 galaxies distributed in 13 clusters, while the zero point is established from 26 calibrators with Cepheid or tip of the red giant branch distances.  The rms scatter in distances found with these galaxies (cluster template and zero point calibrators combined) of 0.42 mag, 21\% in distance, is insignificantly different from the accuracy found with the strongly overlapping $I$ band study.\\

Distance measures derived with this calibration are subject to a small Malmquist bias, requiring the distance modulus correction $\mu^c  = \mu + 0.0065(\mu-31)^2$.  After application of bias and color corrections, a preliminary estimate of the Hubble Constant can be made from the velocities and distances to 7 clusters at $V_{CMB} > 4000$~\kms. Accounting for all error sources, the determination is H$_0 = 74 \pm 5$~\kmsMpc.  The difference between the value determined with this mid-IR analysis compared with the $I$ band value found with the same procedures and a strongly overlapping sample (TC12) is $\Delta {\rm H}_0 = -2$~\kmsMpc, not a formally significant difference.  We reiterate that the great strength of the present calibration is the high confidence in uniformity over the entire sky.  Nevertheless the present sample of only 7 clusters beyond the domain of known extreme peculiar velocities is unsatisfactorily small.  In a subsequent paper \citep{2012ApJ...758L..12S}  the [3.6] band calibration is extended to a calibration of the Type Ia supernova scale, analogous to what has been done at $I$ band \citep{2012ApJ...749..174C}, permitting a determination of H$_0$ at $z \sim 0.1$.\\

\bigskip\bigskip
The data use in this paper are available at the {\it Extragalactic Distance Database}.\footnote{http://edd.ifa.hawaii.edu}  The photometric data are found by selecting the catalog {\it Spitzer [3.6] Band Photometry} and then a galaxy of choice while the HI profiles are found in the catalog {\it All Digital HI}. 
We thank Tom Jarrett, part of the {\it Cosmic Flows with Spitzer} collaboration, for advice with Spitzer photometry and James Shombert for his development of Archangel.  Much of the data used here comes from the {\it Spitzer Space Telescope} archive.  We thank Kartik Sheth for the contribution of his program {\it Spitzer Survey of Stellar Structure in Galaxies}.  NASA through the {\it Spitzer Science Center} provides support for CHP, the {\it Carnegie Hubble Program}, cycle 6 program 61009, and for CFS, {\it Cosmic Flows with Spitzer}, cycle 8 program 80072.  RBT receives support from the US National Science Foundation with award AST-0908846.

\clearpage

\begin{table}[h!]
\caption{Calibrator parameters\tablenotemark{x}}
\label{tbl:online}
\begin{tabular}{|c|c|c|c|c|c|c|c|c|c|c|c|}
\hline
PGC\tablenotemark{a} & Name\tablenotemark{b} & $[3.6]_{ext}$\tablenotemark{c} & $[3.6]_{CHP}$\tablenotemark{d} & $[3.6]_{ave}$\tablenotemark{e} & $I-[3.6]_{ave}$\tablenotemark{f} & b/a\tablenotemark{g} & Inc\tablenotemark{h} & $W_{mx}$\tablenotemark{i} & $W^i_{mx}$\tablenotemark{j} & ${\rm log}(W^i_{mx})$\tablenotemark{k} & Sam\tablenotemark{l}  \\
\hline
   9332 &  NGC0925 &  10.866 &  10.231 & 10.549 & -1.589 &  0.57 &  57 & 194 & 231 & 2.364 & ZerpPt\\
  13179 & NGC1365 &    8.818 &    8.812 &   8.815 &  -0.725 &  0.61 &  54 & 371 & 459 & 2.662 & ZeroPt\\
  13602 & NGC1425 &  10.693 & 10.700 & 10.697 &  -1.197 &  0.46 &  65 & 354 & 391 & 2.592 & ZeroPt\\
  17819 & NGC2090 &  10.477 & 10.287 & 10.382 &  -1.052 &  0.43 &  67 & 277 & 301 & 2.478 & ZeroPt\\
  21396 & NGC2403 &    8.558 &    8.370 &   8.464 &  -1.354 &  0.53 &  60 & 226 & 261 & 2.417 & ZeroPt\\
  23110 & NGC2541 &       /       &  11.949 & 11.949 & -1.189 &  0.49 &  63 & 188 & 211 & 2.325 & ZeroPt\\
  26512 & NGC2841 &    8.644 &   8.644   &   8.644 &  -1.114 &  0.45 &  66 & 592 & 650 & 2.813 & ZeroPt\\
  28120 & NGC2976 &    9.904 &       /        &   9.904 &  -0.924 &  0.53  & 60 & 129 & 149 & 2.173 & ZeroPt\\
  28357 & NGC3021 &  11.693 &       /        & 11.693 &  -0.773 &  0.57  & 57 & 254 & 303 & 2.481 & ZeroPt\\
  30197 & NGC3198 &  10.368 &   10.326 & 10.347 & -1.177 &  0.39  & 70 & 296 & 315 & 2.498 & ZeroPt\\
  30819 & IC2574      &  11.750 &       /      & 11.750   & -1.630  &  0.40  &  69.  &   106   & 113.  &  2.054    & ZeroPt\\
  32007 & NGC3351 &    9.208 &    9.210 &    9.209 &   -0.879  &  0.70  &  47. &   262  &  359. &   2.556    & ZeroPt\\ 
  32207 & NGC3370 &  11.739 &      /       & 11.739  &  -0.889  &  0.56  &  58.  &  264  &  312.   & 2.494    & ZeroPt\\
  34554 & NGC3621 &    8.989 &    9.035  &   9.012  &  -1.002  &  0.45 &   66. &   266  &  292. &   2.465    & ZeroPt\\
  34695 & NGC3627 &    8.314 &    8.254  &   8.284 &   -0.894  &  0.53 &   60.  &  333  &  385.  &  2.585    & ZeroPt\\
  39422 & NGC4244 &  10.333 &      /        & 10.333  &  -1.413  &  0.20  &  90.  &  192  &  192.  &  2.283    & ZeroPt\\
  40692 & NGC4414 &    9.368 &    9.367  &   9.368 &   -0.638  &  0.60 &   55.  &  378   & 463.  &  2.666    & ZeroPt\\
  41812 & NGC4535 &    9.783 &    9.751   &  9.767  &  -0.817 &   0.72  &  45.  &  265  &  374.  &  2.573    & ZeroPt\\
  41823 & NGC4536 &    9.840 &    9.856  &   9.848 &   -0.818 &   0.38  &  71.  &  322  &  341. &   2.533    & ZeroPt\\
  42408 & NGC4605 &  10.161 &      /        & 10.161  &  -0.971  &  0.41 &   69.  &  154  &  165.  &  2.219    & ZeroPt\\
  42510 & NGC4603 &  10.682 &   10.663  &  10.673 &   -0.913 &   0.64 &   52. &   353  &  450.  &  2.653    & ZeroPt\\
  42741 & NGC4639 &  11.250 &   11.255  &  11.253  &  -1.073   & 0.60  &  55.   & 274  &  336.  &  2.526    & ZeroPt\\
  43451 & NGC4725 &    8.922 &    8.893  &   8.908  &  -1.068  &  0.56  &  58.  &  397  &  470.  &  2.672    & ZeroPt\\
  51344 & NGC5584 &  11.763 &   11.819  &  11.791  &  -1.171  &  0.73  &  44.  &  186  &  267.  &  2.426    & ZeroPt\\ 
  69327 & NGC7331 &    8.409 &    8.377   &  8.393  &  -0.873  &  0.44  &  66.  &  501  &  547.  &  2.738    & ZeroPt\\
  73049 & NGC7793 &     9.298 &       /    &   9.298  &  -1.048  &  0.62 &   53.  &  162  &  202.  &  2.306    & ZeroPt\\
  ... & & & & & & & & & & &\\
\hline
\end{tabular}
\tablenotemark{x}{Complete table online }\\
\tablenotemark{a}{PGC number }\\
\tablenotemark{b}{Common Name }\\
\tablenotemark{c}{CFS corrected magnitude, mag }\\
\tablenotemark{d}{CHP corrected magnitude, mag} \\
\tablenotemark{e}{CFS and CHP averaged corrected magnitude, mag}\\
\tablenotemark{f}{Color term $I-[3.6]_{ave}$, mag}\\
\tablenotemark{g}{Axial Ratio }\\
\tablenotemark{h}{Inclination, degrees }\\
\tablenotemark{i}{linewidth not corrected for inclination, \kms }\\
\tablenotemark{j}{linewidth corrected for inclination, \kms }\\
\tablenotemark{k}{Logarithm of the inclination corrected linewidth}\\
\tablenotemark{l}{Sample}
\end{table}

\clearpage

\begin{table}[h!]
\begin{center}
\caption{Properties of the Cluster Fits}
\label{tbl:clfits}
\begin{tabular}{|l|c|c|r|c|c|c|c|c|c|c|c|}
\hline
Cluster\tablenotemark{a} & $V_{CMB}$\tablenotemark{b} & N\tablenotemark{c} & Slope\tablenotemark{d} & ZP\tablenotemark{e} & rms\tablenotemark{f} & ZP$_{color}$\tablenotemark{g} & rms\tablenotemark{h} & bias\tablenotemark{i} & DM\tablenotemark{j} & Dist\tablenotemark{k} & $V/D$\tablenotemark{l} \\
\hline
Virgo & 1410 & 24 & -8.21 $\pm$ 0.71 &10.40 $\pm$  0.14 & 0.67 &  10.49 $\pm$ 0.11 & 0.56 & 0.00 & 30.83 $\pm$ 0.14 &14.7 $\pm$ 0.9 & 96.2 $\pm$ 6.9\\ 
Fornax & 1484 &  15 & -9.39 $\pm$ 0.66 & 10.73 $\pm$ 0.13 & 0.49 & 10.86 $\pm$ 0.12 & 0.47 & 0.00 & 31.20 $\pm$ 0.14 &17.4 $\pm$ 1.2 & 85.4 $\pm$ 6.4\\
U Ma & 1101 & 32 & -9.83 $\pm$ 0.52 & 10.84 $\pm$ 0.10 & 0.55 & 10.94 $\pm$ 0.08 & 0.44 & 0.00 & 31.28 $\pm$ 0.11 & 18.0 $\pm$ 0.9 & 61.1 $\pm$ 4.2\\
Antlia & 3119 & 11 & -10.79 $\pm$ 0.79 & 12.47 $\pm$ 0.07 & 0.23 & 12.44 $\pm$ 0.06 & 0.21 & 0.04 & 32.82 $\pm$ 0.10 & 36.6 $\pm$ 1.7 & 85.1 $\pm$ 4.2\\
Cen30 & 3679 & 11 & -12.69 $\pm$ 1.76 & 12.57 $\pm$0.19 & 0.62 & 12.58 $\pm$ 0.18 & 0.59 & 0.01 & 32.93 $\pm$ 0.20 & 38.5 $\pm$ 3.5 & 95.4 $\pm$ 8.8\\
Pegasus & 3518 & 12 & -8.55 $\pm$ 0.94 & 12.87 $\pm$ 0.13 & 0.44 & 12.89 $\pm$ 0.12 & 0.42 & 0.01 & 33.24 $\pm$ 0.14 & 44.5 $\pm$ 3.0 & 79.1 $\pm$ 5.4\\
Hydra & 4121 & 14 & -10.48 $\pm$ 1.49 & 13.30 $\pm$ 0.14 & 0.53 & 13.35 $\pm$ 0.14 & 0.52 & 0.05 & 33.74 $\pm$ 0.16 & 56.0 $\pm$ 4.2 & 73.6 $\pm$ 5.5\\
Pisces & 4779 & 23 & -10.15 $\pm$ 0.82 & 13.70 $\pm$ 0.10 & 0.47 & 13.69 $\pm$ 0.08 & 0.40 & 0.02 & 34.05 $\pm$ 0.11 & 64.6 $\pm$ 3.4 & 74.0 $\pm$ 3.9\\
Cancer & 4940 & 11 & -11.46 $\pm$ 1.17 & 13.77 $\pm$ 0.12 & 0.39 & 13.78 $\pm$ 0.10 & 0.34 & 0.02 & 34.14 $\pm$ 0.13 & 67.3 $\pm$ 4.0 & 73.4 $\pm$ 4.4\\
Coma & 7194 & 16 & -8.49 $\pm$ 1.10 & 14.44 $\pm$ 0.12 & 0.49 & 14.50 $\pm$ 0.10 & 0.39 & 0.06 & 34.90 $\pm$ 0.13 & 95.4 $\pm$ 5.6 & 75.3 $\pm$ 4.5\\
A400 & 7108 & 7 & -8.03 $\pm$ 1.40 & 14.52 $\pm$ 0.08 & 0.21 & 14.50 $\pm$ 0.08 & 0.21 & 0.10 & 34.94 $\pm$ 0.11 & 97.3 $\pm$ 5.1 & 73.1 $\pm$ 3.8\\ 
A1367 & 6923 & 19 & -9.38 $\pm$ 1.16 & 14.50 $\pm$ 0.11 & 0.47 & 14.50 $\pm$ 0.11 & 0.42 & 0.08 & 34.92 $\pm$ 0.14 & 96.4 $\pm$ 6.0 & 71.8 $\pm$ 4.5\\
A2634/66 & 8381 & 18 & -9.55 $\pm$ 1.38 & 14.83 $\pm$ 0.12 & 0.51 & 14.85 $\pm$ 0.10 & 0.44 & 0.05 & 35.24 $\pm$ 0.13 & 111.7 $\pm$ 6.6 & 75.0 $\pm$ 4.4\\
\hline
\end{tabular}
\end{center}
\tablenotemark{a}{Cluster name} \\
\tablenotemark{b}{Mean velocity of the cluster with respect to the CMB, \kms }\\
\tablenotemark{c}{Number of studied galaxy per cluster }\\
\tablenotemark{d}{Slope of the inverse fit }\\
\tablenotemark{e}{Zero point relative to Virgo's zero point, no color adjustment, mag }\\
\tablenotemark{f}{Scatter, no color adjustment }\\
\tablenotemark{g}{Zero point relative to Virgo's zero point after color adjustment, mag }\\
\tablenotemark{h}{Scatter after color adjustment, mag }\\
\tablenotemark{i}{Bias, mag }\\
\tablenotemark{j}{Bias corrected Distance Modulus, mag }\\
\tablenotemark{k}{Cluster Distance, Mpc }\\
\tablenotemark{l}{Hubble parameter, \kmsMpc}

\end{table}

\begin{table}[h!]
\caption{TFR parameters}
\label{tbl:compare}
\begin{tabular}{|l|c|c|c|c|}
\hline
 Sample                 & Ngal  & Slope  & RMS & Zero Point \\
\hline
$I$ template         &   267 & -8.81$\pm$0.16 & 0.41 &  -- \\
$I$ zero point       &    36  &       --      &   0.36 &  -21.39$\pm$0.07 (Veg) \\
$[3.6]$ template   &  213 & -9.74$\pm$0.22 & 0.49 &  -- \\
$[3.6]$ zero point &    26 &        --     &    0.44 &  -20.34$\pm$0.10 (AB) \\
$M_C$ template  &  213 & -9.13$\pm$0.22 & 0.44 &  -- \\
$M_C$ zero point &   26  &       --     &    0.37 &  -20.34$\pm$0.08 (AB) \\
\hline
\end{tabular}
\end{table}

\begin{table}[h!]
\caption{Comparison with \citet{2012ApJ...749..174C}}
\label{tbl:compIcal}
\begin{tabular}{|c|c|c|c|c|c|}
\hline
Cluster & This Paper\tablenotemark{a} & TC12\tablenotemark{a} & Cluster & This Paper\tablenotemark{a} & TC12\tablenotemark{a}\\
\hline
V~Virgo & 14.7 $\pm$ 0.9 & 15.9 $\pm$ 0.8 & Pi~Pisces & 65 $\pm$ 3 & 64 $\pm$ 2 \\
F~Fornax & 17.4 $\pm$ 1.2 & 17.3 $\pm$ 1.0 & Ca~Cancer & 67 $\pm$ 4 & 65 $\pm$ 3 \\
U~U Ma & 18.0 $\pm$ 0.9 & 17.4 $\pm$ 0.9 & Co~Coma & 95 $\pm$ 6 & 90 $\pm$ 4\\ 
An~Antlia & 37 $\pm$ 2 & 37 $\pm$ 2 & A4~A400 & 97 $\pm$ 5 & 94 $\pm$ 5\\  
Ce~Cen30 & 39 $\pm$ 4 & 38 $\pm$ 3 & A1~A1367 &  96 $\pm$ 6 & 94 $\pm$ 5\\
Pe~Pegasus & 45 $\pm$ 3 & 43 $\pm$ 3 & A2~A2634/66 & 112 $\pm$ 7 & / \\
H~Hydra & 56 $\pm$ 4 & 59 $\pm$ 4 & A2634 & / & 121 $\pm$ 7 \\
\hline
\end{tabular}
\tablenotemark{a}{Distance, Mpc} 
\end{table}

\clearpage

\bibliographystyle{Apj}

\bibliography{bibli2}

\begin{thebibliography}{40}
\expandafter\ifx\csname natexlab\endcsname\relax\def\natexlab#1{#1}\fi

\bibitem[{{Aaronson} {et~al.}(1979){Aaronson}, {Huchra}, \&
  {Mould}}]{1979ApJ...229....1A}
{Aaronson}, M., {Huchra}, J., \& {Mould}, J. 1979, \apj, 229, 1

\bibitem[{{Cardelli} {et~al.}(1989){Cardelli}, {Clayton}, \&
  {Mathis}}]{1989ApJ...345..245C}
{Cardelli}, J.~A., {Clayton}, G.~C., \& {Mathis}, J.~S. 1989, \apj, 345, 245

\bibitem[{{Courtois} \& {Tully}(2012)}]{2012ApJ...749..174C}
{Courtois}, H.~M. \& {Tully}, R.~B. 2012, \apj, 749, 174

\bibitem[{{Courtois} {et~al.}(2009){Courtois}, {Tully}, {Fisher}, {Bonhomme},
  {Zavodny}, \& {Barnes}}]{2009AJ....138.1938C}
{Courtois}, H.~M., {Tully}, R.~B., {Fisher}, J.~R., {Bonhomme}, N., {Zavodny},
  M., \& {Barnes}, A. 2009, \aj, 138, 1938

\bibitem[{{Courtois} {et~al.}(2011{\natexlab{a}}){Courtois}, {Tully}, \&
  {H{\'e}raudeau}}]{2011MNRAS.415.1935C}
{Courtois}, H.~M., {Tully}, R.~B., \& {H{\'e}raudeau}, P. 2011{\natexlab{a}},
  \mnras, 415, 1935

\bibitem[{{Courtois} {et~al.}(2011{\natexlab{b}}){Courtois}, {Tully},
  {Makarov}, {Mitronova}, {Koribalski}, {Karachentsev}, \&
  {Fisher}}]{2011MNRAS.414.2005C}
{Courtois}, H.~M., {Tully}, R.~B., {Makarov}, D.~I., {Mitronova}, S.,
  {Koribalski}, B., {Karachentsev}, I.~D., \& {Fisher}, J.~R.
  2011{\natexlab{b}}, \mnras, 414, 2005

\bibitem[{{Dale} {et~al.}(2005){Dale}, {Bendo}, {Engelbracht}, {Gordon},
  {Regan}, {Armus}, {Cannon}, {Calzetti}, {Draine}, {Helou}, {Joseph},
  {Kennicutt}, {Li}, {Murphy}, {Roussel}, {Walter}, {Hanson}, {Hollenbach},
  {Jarrett}, {Kewley}, {Lamanna}, {Leitherer}, {Meyer}, {Rieke}, {Rieke},
  {Sheth}, {Smith}, \& {Thornley}}]{2005ApJ...633..857D}
{Dale}, D.~A., {Bendo}, G.~J., {Engelbracht}, C.~W., {Gordon}, K.~D., {Regan},
  M.~W., {Armus}, L., {Cannon}, J.~M., {Calzetti}, D., {Draine}, B.~T.,
  {Helou}, G., {Joseph}, R.~D., {Kennicutt}, R.~C., {Li}, A., {Murphy}, E.~J.,
  {Roussel}, H., {Walter}, F., {Hanson}, H.~M., {Hollenbach}, D.~J., {Jarrett},
  T.~H., {Kewley}, L.~J., {Lamanna}, C.~A., {Leitherer}, C., {Meyer}, M.~J.,
  {Rieke}, G.~H., {Rieke}, M.~J., {Sheth}, K., {Smith}, J.~D.~T., \&
  {Thornley}, M.~D. 2005, \apj, 633, 857

\bibitem[{{Dale} {et~al.}(2007){Dale}, {Gil de Paz}, {Gordon}, {Hanson},
  {Armus}, {Bendo}, {Bianchi}, {Block}, {Boissier}, {Boselli}, {Buckalew},
  {Buat}, {Burgarella}, {Calzetti}, {Cannon}, {Engelbracht}, {Helou},
  {Hollenbach}, {Jarrett}, {Kennicutt}, {Leitherer}, {Li}, {Madore}, {Martin},
  {Meyer}, {Murphy}, {Regan}, {Roussel}, {Smith}, {Sosey}, {Thilker}, \&
  {Walter}}]{2007ApJ...655..863D}
{Dale}, D.~A., {Gil de Paz}, A., {Gordon}, K.~D., {Hanson}, H.~M., {Armus}, L.,
  {Bendo}, G.~J., {Bianchi}, L., {Block}, M., {Boissier}, S., {Boselli}, A.,
  {Buckalew}, B.~A., {Buat}, V., {Burgarella}, D., {Calzetti}, D., {Cannon},
  J.~M., {Engelbracht}, C.~W., {Helou}, G., {Hollenbach}, D.~J., {Jarrett},
  T.~H., {Kennicutt}, R.~C., {Leitherer}, C., {Li}, A., {Madore}, B.~F.,
  {Martin}, D.~C., {Meyer}, M.~J., {Murphy}, E.~J., {Regan}, M.~W., {Roussel},
  H., {Smith}, J.~D.~T., {Sosey}, M.~L., {Thilker}, D.~A., \& {Walter}, F.
  2007, \apj, 655, 863

\bibitem[{{Eddington}(1913)}]{1913MNRAS..73..359E}
{Eddington}, A.~S. 1913, \mnras, 73, 359

\bibitem[{{Fazio} {et~al.}(2004){Fazio}, {Hora}, {Allen}, {Ashby}, {Barmby},
  {Deutsch}, {Huang}, {Kleiner}, {Marengo}, {Megeath}, {Melnick}, {Pahre},
  {Patten}, {Polizotti}, {Smith}, {Taylor}, {Wang}, {Willner}, {Hoffmann},
  {Pipher}, {Forrest}, {McMurty}, {McCreight}, {McKelvey}, {McMurray}, {Koch},
  {Moseley}, {Arendt}, {Mentzell}, {Marx}, {Losch}, {Mayman}, {Eichhorn},
  {Krebs}, {Jhabvala}, {Gezari}, {Fixsen}, {Flores}, {Shakoorzadeh}, {Jungo},
  {Hakun}, {Workman}, {Karpati}, {Kichak}, {Whitley}, {Mann}, {Tollestrup},
  {Eisenhardt}, {Stern}, {Gorjian}, {Bhattacharya}, {Carey}, {Nelson},
  {Glaccum}, {Lacy}, {Lowrance}, {Laine}, {Reach}, {Stauffer}, {Surace},
  {Wilson}, {Wright}, {Hoffman}, {Domingo}, \& {Cohen}}]{2004ApJS..154...10F}
{Fazio}, G.~G., {Hora}, J.~L., {Allen}, L.~E., {Ashby}, M.~L.~N., {Barmby}, P.,
  {Deutsch}, L.~K., {Huang}, J.-S., {Kleiner}, S., {Marengo}, M., {Megeath},
  S.~T., {Melnick}, G.~J., {Pahre}, M.~A., {Patten}, B.~M., {Polizotti}, J.,
  {Smith}, H.~A., {Taylor}, R.~S., {Wang}, Z., {Willner}, S.~P., {Hoffmann},
  W.~F., {Pipher}, J.~L., {Forrest}, W.~J., {McMurty}, C.~W., {McCreight},
  C.~R., {McKelvey}, M.~E., {McMurray}, R.~E., {Koch}, D.~G., {Moseley}, S.~H.,
  {Arendt}, R.~G., {Mentzell}, J.~E., {Marx}, C.~T., {Losch}, P., {Mayman}, P.,
  {Eichhorn}, W., {Krebs}, D., {Jhabvala}, M., {Gezari}, D.~Y., {Fixsen},
  D.~J., {Flores}, J., {Shakoorzadeh}, K., {Jungo}, R., {Hakun}, C., {Workman},
  L., {Karpati}, G., {Kichak}, R., {Whitley}, R., {Mann}, S., {Tollestrup},
  E.~V., {Eisenhardt}, P., {Stern}, D., {Gorjian}, V., {Bhattacharya}, B.,
  {Carey}, S., {Nelson}, B.~O., {Glaccum}, W.~J., {Lacy}, M., {Lowrance},
  P.~J., {Laine}, S., {Reach}, W.~T., {Stauffer}, J.~A., {Surace}, J.~A.,
  {Wilson}, G., {Wright}, E.~L., {Hoffman}, A., {Domingo}, G., \& {Cohen}, M.
  2004, \apjs, 154, 10

\bibitem[{{Freedman} {et~al.}(2001){Freedman}, {Madore}, {Gibson}, {Ferrarese},
  {Kelson}, {Sakai}, {Mould}, {Kennicutt}, {Ford}, {Graham}, {Huchra},
  {Hughes}, {Illingworth}, {Macri}, \& {Stetson}}]{2001ApJ...553...47F}
{Freedman}, W.~L., {Madore}, B.~F., {Gibson}, B.~K., {Ferrarese}, L., {Kelson},
  D.~D., {Sakai}, S., {Mould}, J.~R., {Kennicutt}, Jr., R.~C., {Ford}, H.~C.,
  {Graham}, J.~A., {Huchra}, J.~P., {Hughes}, S.~M.~G., {Illingworth}, G.~D.,
  {Macri}, L.~M., \& {Stetson}, P.~B. 2001, \apj, 553, 47

\bibitem[{{Freedman} {et~al.}(2012){Freedman}, {Madore}, {Scowcroft}, {Burns},
  {Monson}, {Persson}, {Seibert}, \& {Rigby}}]{freedman2012}
{Freedman}, W.~L., {Madore}, B.~F., {Scowcroft}, V., {Burns}, C., {Monson}, A.,
  {Persson}, S.~E., {Seibert}, M., \& {Rigby}, J. 2012, ArXiv e-prints

\bibitem[{{Freedman} {et~al.}(2011){Freedman}, {Madore}, {Scowcroft}, {Monson},
  {Persson}, {Seibert}, {Rigby}, {Sturch}, \& {Stetson}}]{2011AJ....142..192F}
{Freedman}, W.~L., {Madore}, B.~F., {Scowcroft}, V., {Monson}, A., {Persson},
  S.~E., {Seibert}, M., {Rigby}, J.~R., {Sturch}, L., \& {Stetson}, P. 2011,
  \aj, 142, 192

\bibitem[{{Giovanelli} {et~al.}(1997{\natexlab{a}}){Giovanelli}, {Haynes},
  {Herter}, {Vogt}, {da Costa}, {Freudling}, {Salzer}, \&
  {Wegner}}]{1997AJ....113...53G}
{Giovanelli}, R., {Haynes}, M.~P., {Herter}, T., {Vogt}, N.~P., {da Costa},
  L.~N., {Freudling}, W., {Salzer}, J.~J., \& {Wegner}, G. 1997{\natexlab{a}},
  \aj, 113, 53

\bibitem[{{Giovanelli} {et~al.}(1997{\natexlab{b}}){Giovanelli}, {Haynes},
  {Herter}, {Vogt}, {Wegner}, {Salzer}, {da Costa}, \&
  {Freudling}}]{1997AJ....113...22G}
{Giovanelli}, R., {Haynes}, M.~P., {Herter}, T., {Vogt}, N.~P., {Wegner}, G.,
  {Salzer}, J.~J., {da Costa}, L.~N., \& {Freudling}, W. 1997{\natexlab{b}},
  \aj, 113, 22

\bibitem[{{Giovanelli} {et~al.}(1995){Giovanelli}, {Haynes}, {Salzer},
  {Wegner}, {da Costa}, \& {Freudling}}]{1995AJ....110.1059G}
{Giovanelli}, R., {Haynes}, M.~P., {Salzer}, J.~J., {Wegner}, G., {da Costa},
  L.~N., \& {Freudling}, W. 1995, \aj, 110, 1059

\bibitem[{{Huang} {et~al.}(2007){Huang}, {Ashby}, {Barmby}, {Brodwin}, {Brown},
  {Caldwell}, {Cool}, {Eisenhardt}, {Eisenstein}, {Fazio}, {Le Floc'h},
  {Green}, {Kochanek}, {Lu}, {Pahre}, {Rigopoulou}, {Rosenberg}, {Smith},
  {Wang}, {Willmer}, \& {Willner}}]{2007ApJ...664..840H}
{Huang}, J.-S., {Ashby}, M.~L.~N., {Barmby}, P., {Brodwin}, M., {Brown},
  M.~J.~I., {Caldwell}, N., {Cool}, R.~J., {Eisenhardt}, P., {Eisenstein}, D.,
  {Fazio}, G.~G., {Le Floc'h}, E., {Green}, P., {Kochanek}, C.~S., {Lu}, N.,
  {Pahre}, M.~A., {Rigopoulou}, D., {Rosenberg}, J.~L., {Smith}, H.~A., {Wang},
  Z., {Willmer}, C.~N.~A., \& {Willner}, S.~P. 2007, \apj, 664, 840

\bibitem[{{Karachentsev} {et~al.}(2002){Karachentsev}, {Mitronova},
  {Karachentseva}, {Kudrya}, \& {Jarrett}}]{2002A&A...396..431K}
{Karachentsev}, I.~D., {Mitronova}, S.~N., {Karachentseva}, V.~E., {Kudrya},
  Y.~N., \& {Jarrett}, T.~H. 2002, \aap, 396, 431

\bibitem[{{Masters} {et~al.}(2006){Masters}, {Springob}, {Haynes}, \&
  {Giovanelli}}]{2006ApJ...653..861M}
{Masters}, K.~L., {Springob}, C.~M., {Haynes}, M.~P., \& {Giovanelli}, R. 2006,
  \apj, 653, 861

\bibitem[{{Meidt} {et~al.}(2012){Meidt}, {Schinnerer}, {Knapen}, {Bosma},
  {Athanassoula}, {Sheth}, {Buta}, {Zaritsky}, {Laurikainen}, {Elmegreen},
  {Elmegreen}, {Gadotti}, {Salo}, {Regan}, {Ho}, {Madore}, {Hinz}, {Skibba},
  {Gil de Paz}, {Mu{\~n}oz-Mateos}, {Men{\'e}ndez-Delmestre}, {Seibert}, {Kim},
  {Mizusawa}, {Laine}, \& {Comer{\'o}n}}]{2012ApJ...744...17M}
{Meidt}, S.~E., {Schinnerer}, E., {Knapen}, J.~H., {Bosma}, A., {Athanassoula},
  E., {Sheth}, K., {Buta}, R.~J., {Zaritsky}, D., {Laurikainen}, E.,
  {Elmegreen}, D., {Elmegreen}, B.~G., {Gadotti}, D.~A., {Salo}, H., {Regan},
  M., {Ho}, L.~C., {Madore}, B.~F., {Hinz}, J.~L., {Skibba}, R.~A., {Gil de
  Paz}, A., {Mu{\~n}oz-Mateos}, J.-C., {Men{\'e}ndez-Delmestre}, K., {Seibert},
  M., {Kim}, T., {Mizusawa}, T., {Laine}, J., \& {Comer{\'o}n}, S. 2012, \apj,
  744, 17

\bibitem[{{Oke} \& {Sandage}(1968)}]{1968ApJ...154...21O}
{Oke}, J.~B. \& {Sandage}, A. 1968, \apj, 154, 21

\bibitem[{{Reach} {et~al.}(2005){Reach}, {Megeath}, {Cohen}, {Hora}, {Carey},
  {Surace}, {Willner}, {Barmby}, {Wilson}, {Glaccum}, {Lowrance}, {Marengo}, \&
  {Fazio}}]{2005PASP..117..978R}
{Reach}, W.~T., {Megeath}, S.~T., {Cohen}, M., {Hora}, J., {Carey}, S.,
  {Surace}, J., {Willner}, S.~P., {Barmby}, P., {Wilson}, G., {Glaccum}, W.,
  {Lowrance}, P., {Marengo}, M., \& {Fazio}, G.~G. 2005, \pasp, 117, 978

\bibitem[{{Riess} {et~al.}(2011){Riess}, {Macri}, {Casertano}, {Lampeitl},
  {Ferguson}, {Filippenko}, {Jha}, {Li}, \& {Chornock}}]{2011ApJ...730..119R}
{Riess}, A.~G., {Macri}, L., {Casertano}, S., {Lampeitl}, H., {Ferguson},
  H.~C., {Filippenko}, A.~V., {Jha}, S.~W., {Li}, W., \& {Chornock}, R. 2011,
  \apj, 730, 119

\bibitem[{{Rizzi} {et~al.}(2007){Rizzi}, {Tully}, {Makarov}, {Makarova},
  {Dolphin}, {Sakai}, \& {Shaya}}]{2007ApJ...661..815R}
{Rizzi}, L., {Tully}, R.~B., {Makarov}, D., {Makarova}, L., {Dolphin}, A.~E.,
  {Sakai}, S., \& {Shaya}, E.~J. 2007, \apj, 661, 815

\bibitem[{{Rubin} {et~al.}(1985){Rubin}, {Burstein}, {Ford}, \&
  {Thonnard}}]{1985ApJ...289...81R}
{Rubin}, V.~C., {Burstein}, D., {Ford}, Jr., W.~K., \& {Thonnard}, N. 1985,
  \apj, 289, 81

\bibitem[{{Schechter}(1976)}]{1976ApJ...203..297S}
{Schechter}, P. 1976, \apj, 203, 297

\bibitem[{{Schlegel} {et~al.}(1998){Schlegel}, {Finkbeiner}, \&
  {Davis}}]{1998ApJ...500..525S}
{Schlegel}, D.~J., {Finkbeiner}, D.~P., \& {Davis}, M. 1998, \apj, 500, 525

\bibitem[{{Schombert}(2007)}]{2007astro.ph..3646S}
{Schombert}, J. 2007, ArXiv Astrophysics e-prints

\bibitem[{{Sheth} {et~al.}(2010){Sheth}, {Regan}, {Hinz}, {Gil de Paz},
  {Men{\'e}ndez-Delmestre}, {Mu{\~n}oz-Mateos}, {Seibert}, {Kim},
  {Laurikainen}, {Salo}, {Gadotti}, {Laine}, {Mizusawa}, {Armus},
  {Athanassoula}, {Bosma}, {Buta}, {Capak}, {Jarrett}, {Elmegreen},
  {Elmegreen}, {Knapen}, {Koda}, {Helou}, {Ho}, {Madore}, {Masters},
  {Mobasher}, {Ogle}, {Peng}, {Schinnerer}, {Surace}, {Zaritsky},
  {Comer{\'o}n}, {de Swardt}, {Meidt}, {Kasliwal}, \&
  {Aravena}}]{2010PASP..122.1397S}
{Sheth}, K., {Regan}, M., {Hinz}, J.~L., {Gil de Paz}, A.,
  {Men{\'e}ndez-Delmestre}, K., {Mu{\~n}oz-Mateos}, J.-C., {Seibert}, M.,
  {Kim}, T., {Laurikainen}, E., {Salo}, H., {Gadotti}, D.~A., {Laine}, J.,
  {Mizusawa}, T., {Armus}, L., {Athanassoula}, E., {Bosma}, A., {Buta}, R.~J.,
  {Capak}, P., {Jarrett}, T.~H., {Elmegreen}, D.~M., {Elmegreen}, B.~G.,
  {Knapen}, J.~H., {Koda}, J., {Helou}, G., {Ho}, L.~C., {Madore}, B.~F.,
  {Masters}, K.~L., {Mobasher}, B., {Ogle}, P., {Peng}, C.~Y., {Schinnerer},
  E., {Surace}, J.~A., {Zaritsky}, D., {Comer{\'o}n}, S., {de Swardt}, B.,
  {Meidt}, S.~E., {Kasliwal}, M., \& {Aravena}, M. 2010, \pasp, 122, 1397

\bibitem[{{Sorce} {et~al.}(2012{\natexlab{a}}){Sorce}, {Courtois}, \&
  {Tully}}]{2012AJ....144..133S}
{Sorce}, J.~G., {Courtois}, H.~M., \& {Tully}, R.~B. 2012{\natexlab{a}}, \aj,
  144, 133

\bibitem[{{Sorce} {et~al.}(2012{\natexlab{b}}){Sorce}, {Tully}, \&
  {Courtois}}]{2012ApJ...758L..12S}
{Sorce}, J.~G., {Tully}, R.~B., \& {Courtois}, H.~M. 2012{\natexlab{b}}, \apjl,
  758, L12

\bibitem[{{Tully} \& {Courtois}(2012)}]{2012ApJ...749...78T}
{Tully}, R.~B. \& {Courtois}, H.~M. 2012, \apj, 749, 78

\bibitem[{{Tully} \& {Fisher}(1977)}]{1977A&A....54..661T}
{Tully}, R.~B. \& {Fisher}, J.~R. 1977, \aap, 54, 661

\bibitem[{{Tully} {et~al.}(1982){Tully}, {Mould}, \&
  {Aaronson}}]{1982ApJ...257..527T}
{Tully}, R.~B., {Mould}, J.~R., \& {Aaronson}, M. 1982, \apj, 257, 527

\bibitem[{{Tully} \& {Pierce}(2000)}]{2000ApJ...533..744T}
{Tully}, R.~B. \& {Pierce}, M.~J. 2000, \apj, 533, 744

\bibitem[{{Tully} {et~al.}(1998){Tully}, {Pierce}, {Huang}, {Saunders},
  {Verheijen}, \& {Witchalls}}]{1998AJ....115.2264T}
{Tully}, R.~B., {Pierce}, M.~J., {Huang}, J.-S., {Saunders}, W., {Verheijen},
  M.~A.~W., \& {Witchalls}, P.~L. 1998, \aj, 115, 2264

\bibitem[{{Tully} {et~al.}(2008){Tully}, {Shaya}, {Karachentsev}, {Courtois},
  {Kocevski}, {Rizzi}, \& {Peel}}]{2008ApJ...676..184T}
{Tully}, R.~B., {Shaya}, E.~J., {Karachentsev}, I.~D., {Courtois}, H.~M.,
  {Kocevski}, D.~D., {Rizzi}, L., \& {Peel}, A. 2008, \apj, 676, 184

\bibitem[{{Werner} {et~al.}(2004){Werner}, {Roellig}, {Low}, {Rieke}, {Rieke},
  {Hoffmann}, {Young}, {Houck}, {Brandl}, {Fazio}, {Hora}, {Gehrz}, {Helou},
  {Soifer}, {Stauffer}, {Keene}, {Eisenhardt}, {Gallagher}, {Gautier}, {Irace},
  {Lawrence}, {Simmons}, {Van Cleve}, {Jura}, {Wright}, \&
  {Cruikshank}}]{2004ApJS..154....1W}
{Werner}, M.~W., {Roellig}, T.~L., {Low}, F.~J., {Rieke}, G.~H., {Rieke}, M.,
  {Hoffmann}, W.~F., {Young}, E., {Houck}, J.~R., {Brandl}, B., {Fazio}, G.~G.,
  {Hora}, J.~L., {Gehrz}, R.~D., {Helou}, G., {Soifer}, B.~T., {Stauffer}, J.,
  {Keene}, J., {Eisenhardt}, P., {Gallagher}, D., {Gautier}, T.~N., {Irace},
  W., {Lawrence}, C.~R., {Simmons}, L., {Van Cleve}, J.~E., {Jura}, M.,
  {Wright}, E.~L., \& {Cruikshank}, D.~P. 2004, \apjs, 154, 1

\bibitem[{{Willick}(1994)}]{1994ApJS...92....1W}
{Willick}, J.~A. 1994, \apjs, 92, 1

\bibitem[{{Willick} {et~al.}(1995){Willick}, {Courteau}, {Faber}, {Burstein},
  \& {Dekel}}]{1995ApJ...446...12W}
{Willick}, J.~A., {Courteau}, S., {Faber}, S.~M., {Burstein}, D., \& {Dekel},
  A. 1995, \apj, 446, 12

\end{thebibliography}

\end{document}